\documentclass[preprint,12pt,3p]{elsarticle}

\usepackage{amssymb}

\usepackage{hyperref}
\usepackage{comment}
\usepackage{multirow}
\usepackage{marginnote}
\usepackage{graphicx}
\usepackage{caption}
\usepackage{subcaption}
\usepackage{tabularx}

\renewcommand*{\today}{October 12, 2017}

\newif\ifremark
\long\def\remark#1{
\ifremark%
        \begingroup%
        \dimen0=\columnwidth
        \advance\dimen0 by -1in%
        \setbox0=\hbox{\parbox[b]{\dimen0}{\protect\em #1}}
        \dimen1=\ht0\advance\dimen1 by 2pt%
        \dimen2=\dp0\advance\dimen2 by 2pt%
        \vskip 0.25pt%
        \hbox to \columnwidth{%
                \vrule height\dimen1 width 3pt depth\dimen2%
                \hss\copy0\hss%
                \vrule height\dimen1 width 3pt depth\dimen2%
        }%
        \endgroup%
\fi}

\remarkfalse

\begin{document}

\begin{frontmatter}

\title{Power-Performance Tradeoffs in Data Center Servers:\\DVFS, CPU pinning, Horizontal, and Vertical Scaling}


\author[label1]{Jakub Krzywda\corref{cor1}}
\address[label1]{Department of Computing Science, Ume\r{a} University, SE-901 87 Ume\r{a}, Sweden}

\cortext[cor1]{Corresponding author}

\ead{jakub@cs.umu.se}
\ead[url]{www.cs.umu.se/~jakub}

\author[label1]{Ahmed Ali-Eldin}
\ead{ahmeda@cs.umu.se}

\author[label2]{Trevor E. Carlson\corref{cor2}}
\address[label2]{School of Computing, National University of Singapore, 13 Computing Drive, Singapore 117417}
\ead{tcarlson@comp.nus.edu.sg}

\cortext[cor2]{This work was completed while Trevor E. Carlson was at Uppsala University.}

\author[label1]{Per-Olov~\"{O}stberg}
\ead{p-o@cs.umu.se}

\author[label1]{Erik~Elmroth}
\ead{elmroth@cs.umu.se}

\begin{abstract}
Dynamic Voltage and Frequency Scaling (DVFS), CPU pinning, horizontal, and vertical scaling, are four techniques that have been proposed as actuators to control the performance and energy consumption on data center servers.
This work investigates the utility of these four actuators, and quantifies the power-performance tradeoffs associated with them.
Using replicas of the German Wikipedia running on our local testbed, we perform a set of experiments to quantify the influence of DVFS, vertical and horizontal scaling, and CPU pinning on end-to-end response time (average and tail), throughput, and power consumption with different workloads.
Results of the experiments show that DVFS rarely reduces the power consumption of underloaded servers by more than 5\%, but it can be used to limit the maximal power consumption of a saturated server by up to 20\% (at a cost of performance degradation).
CPU pinning reduces the power consumption of underloaded server (by up to 7\%) at the cost of performance degradation, which can be limited by choosing an appropriate CPU pinning scheme.
Horizontal and vertical scaling improves both the average and tail response time, but the improvement is not proportional to the amount of resources added.
The load balancing strategy has a big impact on the tail response time of horizontally scaled applications.
\end{abstract}

\begin{keyword}
Power-performance tradeoffs \sep Dynamic Voltage and Frequency Scaling (DVFS) \sep CPU pinning \sep horizontal scaling \sep vertical scaling
\end{keyword}

\end{frontmatter}


\section{Introduction}
Reducing the power consumption of data centers has recently become a major challenge with many efforts from governments and companies around the world.\footnote{For example, the US government, \url{http://datacenters.lbl.gov/}; and the European Union, \url{http://iet.jrc.ec.europa.eu/energyefficiency/ict-codes-conduct/data-centres-energy-efficiency/}.} 
It is estimated that the power consumption of data centers accounts for 1.4\% of the total world consumption~\cite{van2014trends} with some reports suggesting that the total data center energy consumption might rise to up to 8\% of the total consumption by 2020~\cite{koomey2011growth}. 
Some studies found that 56\% of the power consumption of the data center is consumed by the servers, 30\% for cooling,  8\% for power conditioning, and 5\% by the networks~\cite{pelley2009understanding}.
Numerous studies have discussed how to improve data center power consumption by, e.g., improving server resource management~\cite{barroso2013datacenter}, using new server hardware~\cite{kalla2010power7}, and optimizing cooling~\cite{ahuja2012datacenter,zhang2012testore}.
In this paper we focus on the most power consuming element of data centers -- the physical servers.
\remark{Cite numbers from more surveys instead of just two single one -- P-O}

We notice that the energy proportionality has not been achieved yet and an idle physical server consumes a significant amount of power~\cite{barroso2007case, meisner2011power}.
Therefore, two main approaches of physical server resource management has been proposed in the literature to improve the energy efficiency of data center servers: server consolidation and server throttling.

Server consolidation reduces the number of physical servers needed to host a workload through a collocation of applications.
Thanks to that reduction, some servers can be powered down and the rest run at the high utilization level, which is more energy efficient than operating at the lower levels.
However, the server consolidation in a data center with dynamic workloads involves migrations of virtual machines between physical servers in order to reconsolidate workload on a minimal subset of physical servers.
Virtual machine migrations are costly -- they increase the power consumption of physical servers involved in the operation \cite{huang2011power, jeong2013analysis, liu2011performance}, negatively impact the performance of migrated applications \cite{liu2011performance, voorsluys2009cost}, and increase the utilization of physical resources during the migration \cite{jeong2013analysis}.
Moreover, in some cases powering down physical servers is not possible because hosted applications may need resources of a whole cluster to work properly (e.g., memory in case of Google search \cite{lo2014towards}).
Therefore, the workload consolidation can not always be used to effectively improve the energy efficiency of a data center.

Server throttling, 
e.g.,
lowering the frequency and voltage of CPU, reduces the power consumption of an individual server at the cost of computational performance.
It is an approach orthogonal to server consolidation and to some extent can be used in conjunction with it.
This work focuses on the power-performance tradeoffs when using state-of-the-art techniques of server throttling suggested in the literature to reduce the power consumption of data center servers.
We analyse four actuators used to optimize data center servers, namely: Dynamic Voltage and Frequency Scaling (DVFS), which changes the operating frequency and voltage of CPUs; CPU pinning, which defines the set of CPU cores that each thread can run on; vertical and horizontal scaling, which change the amount of resources assigned to a virtual machine and the number of virtual machines, respectively.
We consider three orthogonal dimensions (independent factors) when studying the power-performance tradoffs: a) four actuators, b) different hardware architectures, and c) different workload intensities and distributions.

Savings in the power consumption, thanks to server consolidation or server throttling, do not come for free.
Collocation of applications and reduction of computational capabilities leads to application performance degradation~\cite{nathuji2010qclouds, wang2013impact}.
The response time and throughput, often used as performance indicators, are affected especially when the server becomes saturated.

To analyse the power-performance tradeoffs we perform a set of experiments on a real testbed (described in detail in Section~\ref{sec:testbed}).\footnote{All scripts, virtual machine images, and results are available at \url{http://www8.cs.umu.se/wp/jakub/reproduce/power-performance-tradeoffs/}}
We measure the power consumed by the physical servers and the performance of the applications (response time and throughput) under various configurations.
Then, we analyse the influence of DVFS (Section~\ref{sec:dvfs}); CPU pinning (Section~\ref{sec:cpu-pinning}); as well as horizontal and vertical scaling (Section~\ref{sec:scaling}) on the power-performance tradeoffs.

Our main findings are:
\begin{itemize}
    \item the impact of DVFS on the power consumption of underloaded servers is limited by the CPU idle states (Figure~\ref{fig:dvfs-memcached}h and Table~\ref{tbl:active-and-idle-states-amd-evenly-spread}),
    \item for some request arrival patterns, e.g., bursty arrival pattern, reducing the CPU frequency of underloaded server increases the power consumption (Figure~\ref{fig:dvfs-intel}g),
    \item consolidation of virtual CPUs using CPU pinning reduces the power consumption (Figure~\ref{fig:wiki_horizontal}) at a cost of performance degradation,
    \item the application performance degradation due to consolidation of virtual CPUs can be limited by choosing appropriate CPU pinning scheme (Figure~\ref{fig:memcached_cpu_pinning}),
    \item combining horizontal and vertical scaling with consolidation of virtual CPUs can reduce the power consumption at high resource utilization levels (Figure~\ref{fig:memcached_vertical_horizontal}j,l),
    \item the ability of horizontal and vertical scaling to improve the response time and the throughput is limited when physical servers are highly loaded, but it can be extended by CPU pinning (Figure~\ref{fig:memcached_vertical_horizontal}f,h),
    \item the load balancing strategy has a big impact on the tail response time of horizontally scaled applications (Figure~\ref{fig:horizontal_scaling_load_balancer}c).
\end{itemize}

These findings can be used to control the power budget of a data center and the performance of hosted applications by adjusting the configuration of physical servers (DVFS), virtual machines (horizontal and vertical scaling), and mapping between both (CPU pinning).
Based on these findings, we prepare a set of recommendations for using these actuators, taking into account the power-performance tradeoffs (Table~\ref{tbl:conclusions}).


\section{Related work}
\label{sec:related-work}

DVFS has been used to reduce the power consumption of sever systems by reducing the operating voltage and frequency of the CPU~\cite{aroca2014measurement,tesfatsion2014combined,von2009power}.
DVFS has been studied extensively in the literature with some studies suggesting that DVFS can considerably improve power consumption~\cite{fan2007power,von2009power}, while others show that for some applications the usage of DVFS can increase the overall energy consumption~\cite{le2010dynamic}, or that there is at least a need to do a full system analysis before deciding if DVFS is useful for a workload~\cite{dargie2012analysis, LeSueur:2011:SDS:2002181.2002197}.
The limitations of DVFS, e.g., being coarse grained with only few possible settings, have also been discussed~\cite{lo2014towards}.

CPU pinning has only recently been proposed as one possible way to save energy in data center servers.
Podzimek et al.~\cite{podzimek2015analyzing} analyse how CPU pinning impacts the energy efficiency and the performance interference of two colocated workloads.
They conduct experiments on a real testbed -- an Intel Xeon server (Sandy Bridge), using workloads targeting the JVM (from DaCapo and ScalaBench benchmark suites).
Throughput is used to quantify the performance of applications. Previous work also looked at the performance of databases when CPU pinning is used~\cite{song2009utility}.
Min et al. \cite{min2014development} achieve a speed-up of the execution time of multi-threaded applications by dynamically controlling CPU pinning in Xen hypervisor.
Estrada et al. \cite{estrada2014performance} show that static CPU pinning can improve the performance of an application.

A significant body of work has considered vertical scaling for energy and performance control ~\cite{farokhi2015coordinating, kalyvianaki2009self, spinner2014runtime, lakew2014towards}.
Vertical scaling has been considered as a possible technique to reduce tail response times, reduce overall energy consumption, and improve overall Quality of Service (QoS) of a running application~\cite{lakew2015autonomous}.
Also horizontal scaling has been proposed, as a means to adapt to the changes in the workload, and various controllers were developed to decide when to scale out and in~\cite{alieldin2012adaptive, alieldin2012efficient}.

Studies on using DVFS and CPU pinning do not usually consider the average and tail response times in the evaluation. Studies on using virtual machine elasticity do not consider the case of starting a new virtual machine (horizontal elasticity) instead of adding more cores to the running virtual machine (vertical elasticity).
Almost all of the studies use benchmarking applications, with some of them using old benchmarks such as RUBiS~\cite{farokhi2015coordinating}, or applications that are not widely used~\cite{spinner2014runtime}.

\section{Testbed}
\label{sec:testbed}

In this section, we describe the hardware (physical servers and power distribution units) that we use for our experiments, applications that compose the workload, and a workload generator.

\subsection{Hardware}
\label{sec:testbed_servers}
We run our experiments on two types of physical servers: 
\begin{enumerate}
    \item {HP ProLiant DL165G7} servers are equipped with 32 CPU cores (AMD Opteron\texttrademark \quad 6272, 2.1~GHz), 56~GB of RAM, and 4x500~GB SATA disks arranged in RAID~1+0. 
    The CPU has two sockets, each socket has two Non-Uniform Memory Access (NUMA) nodes, and each NUMA node has 8 cores. 
    Moreover, the cores are arranged in pairs and core pairs share a Floating Point Unit and L2 cache.
    CPU cores on a single die share L3 cache. The idle power consumption of the machine is 120 W.
    The cores can operate at five frequencies: $1.4$, $1.5$, $1.7$, $1.9$, and 2.1 GHz.
    
    \item {Dell R530} physical server has 12 CPU cores (Intel\texttrademark Xeon\texttrademark CPU E5-2620 v3, 2.4~GHz) and 64~GB of RAM.
    The storage consists of 3 Intel DC3610 SSDs and 3x Dell 7200rpm nearline-SAS HDD, both arranged in a hardware RAID~5. 
    There are two CPU sockets, each socket has 6 cores (hyperthreading is disabled for our experiments).
    The cores can operate at frequencies between $1.2$ and 3.2~GHz.
    However, frequencies above 2.4 GHz are turbo frequencies, which means that they can be achieved only on a subset of CPU cores, if that does not cause overheating or power over-consumption.
    The idle power consumption of the machine is 78~W.
\end{enumerate}

\remark{TODO Trevor -- anything missing in the physical servers description?}




We monitor the power consumption of physical servers using HP Intelligent Modular PDUs which provide per-power-socket power usage over Simple Network Management Protocol.
The measurements are done every 0.5~s with accuracy within 1~W.

\subsection{Applications}
To impose a load on the physical servers we use the following applications:


\subsubsection{Stress}
\textit{Stress} is a workload generator for POSIX systems that allows to stress CPU cores\footnote{\url{http://people.seas.harvard.edu/~apw/stress/}}.
When used together with \textit{cpulimit}\footnote{\url{http://cpulimit.sourceforge.net/}}, it enables us to set an arbitrary CPU utilization level.
We use \textit{stress} to investigate the influence of DVFS and CPU pinning on the relation between the CPU utilization and the power consumption.

\subsubsection{MediaWiki}
Wikipedia, the free online encyclopedia, is one of the top 10 accessed websites on the Internet~\cite{alexa}.
The website is managed and maintained by the Wikimedia foundation with the help of many open source volunteers.
The Wikimedia foundation has open sourced MediaWiki, a custom-made, free and open-source wiki software platform written in PHP and JavaScript.
MediaWiki requires an opensource LAMP stack or a WIMP stack (or similar installations)
to function, e.g., Linux or Windows as an operating system, Apache or IIS as a web-server, a MySQL or PostgreSQL databases to store the data, and PHP.
This setup is similar to many cloud applications including, e.g., Facebook which uses a modern version of the LAMP stack~\cite{proctor2014high}, and YouTube~\cite{YouTubeArch}.
In our experiments, we replicate the German Wikipedia on our local testbed. 
We choose the German Wikipedia as it is one of the most popular Wikis in terms of number of users, and in terms of number of articles~\cite{WikiStats}. 
We have chosen to run a setup of the MediaWiki software using MySQL database, enhanced with Memcached\footnote{\url{http://memcached.org/}}, a memory-based object store ``used to speed up dynamic Web applications by alleviating database load''~\cite{fitzpatrick2004distributed}.
In all experiments with multiple virtual machines, we use HAProxy\footnote{\url{http://www.haproxy.org/}} running on a separate physical server for load balancing.

We build a KVM base-image with both Mediawiki and a copy of the German Wiki database.
We then use this base-image to spawn new KVM snapshots of the base-image.
This has two major disadvantages.
First, it wastes storage resources since the snapshot files can end up storing the full database.
Second, it does not allow to easily test requests that write to the database. 
Wikipedia in general receives far more read requests to articles than edit or write requests, making the workload much more read heavy.
Therefore, we replicate both the database and the Mediawiki software in each virtual machine, and send only read requests, what simplifies the experiments significantly while still keeping them realistic.

Usage of MediaWiki application enables us to quantify the influence of DVFS, virtual machine scaling and CPU pinning on the performance.
We have identified thresholds on the average and tail response times that separate two different ways of the application behaviour (see for example Figure~\ref{fig:dvfs-memcached}).
Below the threshold the response time is almost steady or grows slowly with the increasing number of requests, but above the threshold the response time increases rapidly 
(by order of seconds with each workload increase).
Also the power consumption differs, under the threshold it increments, while over the threshold it stabilizes on a constant level.
The threshold on the average end-to-end response is about 1~second and on the tail end-to-end response time for 95th percentile is approximately 2~seconds.
If the application is able to respond to a request under the threshold time we will consider it \textbf{underloaded}, otherwise as an \textbf{overloaded}.
When at least on of the resources of a physical server (in most cases CPU) is fully utilized we will call the physical server \textbf{saturated}.

\remark{rename underloaded -- stable?}
\remark{why do we have the content databases stored inside the VM images? what is this representative of? I'm asking this as I would ask as an external reviewer, describe and motivate. Just saying it made experiments easier may be true but it's not the full story - what impact does this have on the experiments (including how)?}
\remark{MediaWiki: missing information about request patterns, caching, type and amount of resources requested, and level of parallelism in request patterns}

\subsection{Workload generator}
\label{sec:workload-generator}

We have implemented a workload generator using Python that is able to produce different workload profiles. The main workload profiles used in our experiments are:
\begin{enumerate}
    \item Constant -- the number of requests generated per second $n$ does not change over the time of experiment, and the experiment lasts for $r$ repetitions of request generation.
    \item Step -- the number of requests generated per second $n$ grows from 0 to a predefined value $m$ by step of $s$ requests and each step is repeated $r$ times. For example from 0 to $m$=140 requests per second with a step of $s$=5 requests and $r$=2 request generation per step, the number of requests generated per each second will look as follows: 0, 0, 5, 5, 10, 10, {\ldots}, 135, 135.
\end{enumerate}

The request generation process has three modes:
\begin{itemize}
    \item Concurrent -- a client generates a batch of $n$ requests in parallel at the beginning of each second,
    \item Evenly spread -- a client generates a single request every $1/n$ of a second.
    \item Poissonian -- the time between requests generation is determined by a Poisson process.
\end{itemize}
The first two generation modes are extreme cases of all possible distributions of $n$ requests over a second.
The distribution of any real application lies somewhere in between these two.
Poisson process is commonly used to model the request arrival time.
Three request generation modes at the workload generator side translate to three types of request arrival pattern at the server side.

To study the behaviour of a fully loaded system,
the total number of not processed requests in the system is limited to 200 (the value established experimentally, further analysis of the influence of this parameter is provided in Section~\ref{sec:pending-requests-analysis}).
The workload generator delays the generation of a new request if the limit is reached.
The next request is generated when the client receives a response from the server or one of the not processed requests is dropped by the client (10~seconds after its generation).
These properties of the workload generator may change slightly (sub-second order) the request arrival pattern when the application is overloaded.
However, they are useful for the analysis of a fully loaded system, since they stabilise the system for a while before it starts to drop a significant number of requests.
Moreover, delaying the generation of a request is realistic, since it reminds the behaviour of a user waiting for a response to the previous request before sending a next one (e.g., clicking a link to another Wikipedia page).

\section{Dynamic Voltage Frequency Scaling}
\label{sec:dvfs}
Scaling the frequency and voltage of a physical CPU is used to change the performance capabilities and power consumption of a physical server~\cite{qian2011server}.
The performance capabilities are derived from the CPU frequency, while the power consumption of a CPU depends on both the frequency and the voltage (with voltage having a bigger impact).
Each CPU frequency is coupled with an appropriate voltage level and a change of CPU frequency causes a simultaneous change of the voltage.
For multi-core processors with a single voltage regulator, such as our AMD processor, the voltage is associated with the highest CPU frequency in use by any core at given time.
From now on, when we write about scaling the CPU frequency we implicitly refer to changing both the frequency and voltage of a CPU.

\remark{first paragraph can be shortened (extreme case: just state the last sentence)}
\remark{This can be shortened in to, e.g.,:
"DVFS changes both the voltage and the frequency of the CPU.  On our systems, each frequency is coupled with a recommended voltage level by the manufacturer. While one can change the frequency and voltage independently, we change  both the frequency and voltage together to the recommended manufacturer value whenever we refer to changing the frequency."}

\remark{TODO Trevor -- is that more technical now? and correct? /Jakub}

\subsection{Power consumption}
First, we analyse an extreme case, when an application uses only CPU resource (using stress application, similar like in~\cite{aroca2014measurement}) to determine the maximal potential benefit of using DVFS in terms of power savings.
We focus solely on the power consumption of a physical server.


We quantify the influence of the number of fully utilized cores and CPU frequency on the power consumption of a physical server with AMD processors.
We test five CPU utilization levels: 
all cores idle, 
4 cores fully utilized and the rest idle, 
8 cores fully utilized and the rest idle, 
16 cores fully utilized and the rest idle, 
and 32 cores fully utilized;
at five available CPU frequencies: $1.4$, $1.5$, $1.7$, $1.9$, and $2.1$~GHz.
For each of 25 settings (pairs of CPU utilization level and CPU frequency) we monitor the power consumption of a physical server.

\begin{figure}
\centering
\includegraphics[width=0.6\textwidth, trim=0 0 20 30, clip=true]{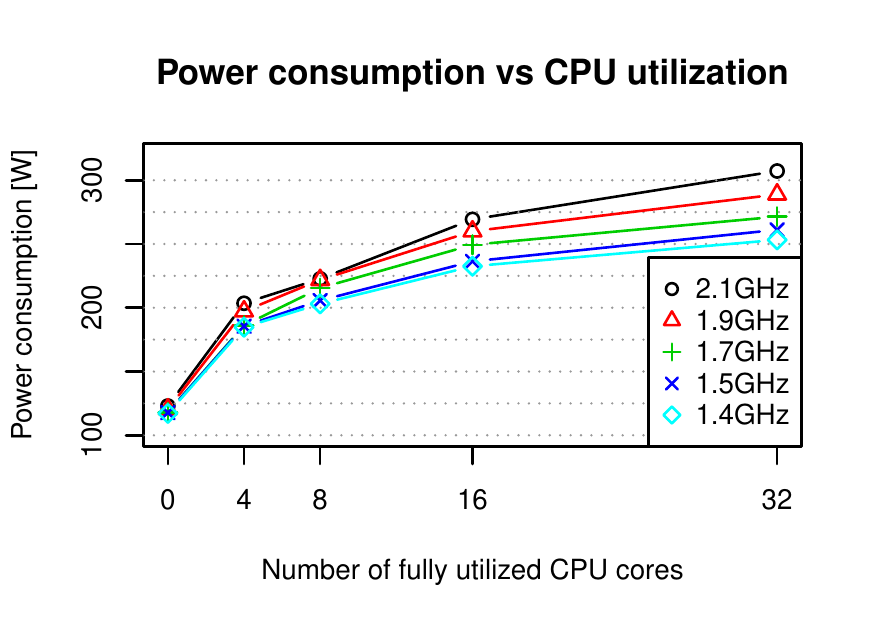}
\caption{The difference in power consumption between various CPU frequencies grows with the number of fully utilized cores.}
\label{fig:power-consumption-dfvs}
\end{figure}

Figure~\ref{fig:power-consumption-dfvs} shows the average values of the power consumption calculated from more than a thousand measurements taken over one hour for each setting.
The difference in power consumption among available CPU frequency/voltage pairs
grows with the increase of the number of utilized CPU cores.
An idle physical server running with the lowest frequency (1.4~GHz) consumes approximately 95\% of the energy consumed at the highest frequency (2.1~GHz).
For a fully utilized physical server the difference is bigger, running a physical server at the frequency of 1.4~GHz saves around 18\% of energy when compared to a physical server operating at the frequency of 2.1~GHz.



\subsection{Power-performance tradeoff}

Until now, we have analyzed the impact of DVFS on the power consumption of a physical server.
However, DVFS has an influence also on the performance of hosted applications, so it is important to analyze both aspects together in order to understand the power-performance tradeoffs.
\remark{I would cut this whole text. It does not say anything. I would start the section by saying "While some  previous work has only considered the power consumption of a server when DVFS is used \ cite{ACMeEnergy}, we perform a set of experiments quantifying..."}
We perform a set of experiments quantifying the influence of workload characteristics on the performance of a MediaWiki application hosted on a physical servers that operate on various CPU frequencies.

\subsubsection{Experiment description}
We run MediaWiki with Memcached in a single virtual machine with 16 physical CPU cores and 20 GB of RAM assigned.
The application is exposed to a step workload that starts at 0 requests per second and continues up to 140 requests per second, with a step of 5 requests, and 10 seconds at each step.
The experiment is repeated on two types of physical servers at various CPU frequencies ($1.4$ and $2.1$~GHz for the AMD processor; and $1.2$, $2.2$ and $3.2$~GHz for the Intel processor), with three types of request arrival pattern (see Section~\ref{sec:workload-generator}).

We are interested in the following metrics:
\begin{itemize}
    \item \textbf{Average response time.} We measure the end-to-end response time at the workload generator (the client side).
    As the average response time we consider the arithmetic mean of all successfully processed requests.
    \item \textbf{Tail response time.} As the tail response time we consider the 95th percentile of all successfully processed requests.
    \item \textbf{Maximal throughput.} As the maximal throughput we consider the maximal number of requests that can be server successfully with the average (or tail) response time under the thresholds.
    \item \textbf{Power consumption.} As the power consumption we consider the total power consumption of a physical server measured at a power socket.
\end{itemize}

\begin{figure*}[t]
    \centering
    \includegraphics[width=\textwidth, trim={0 0 0 0},clip]{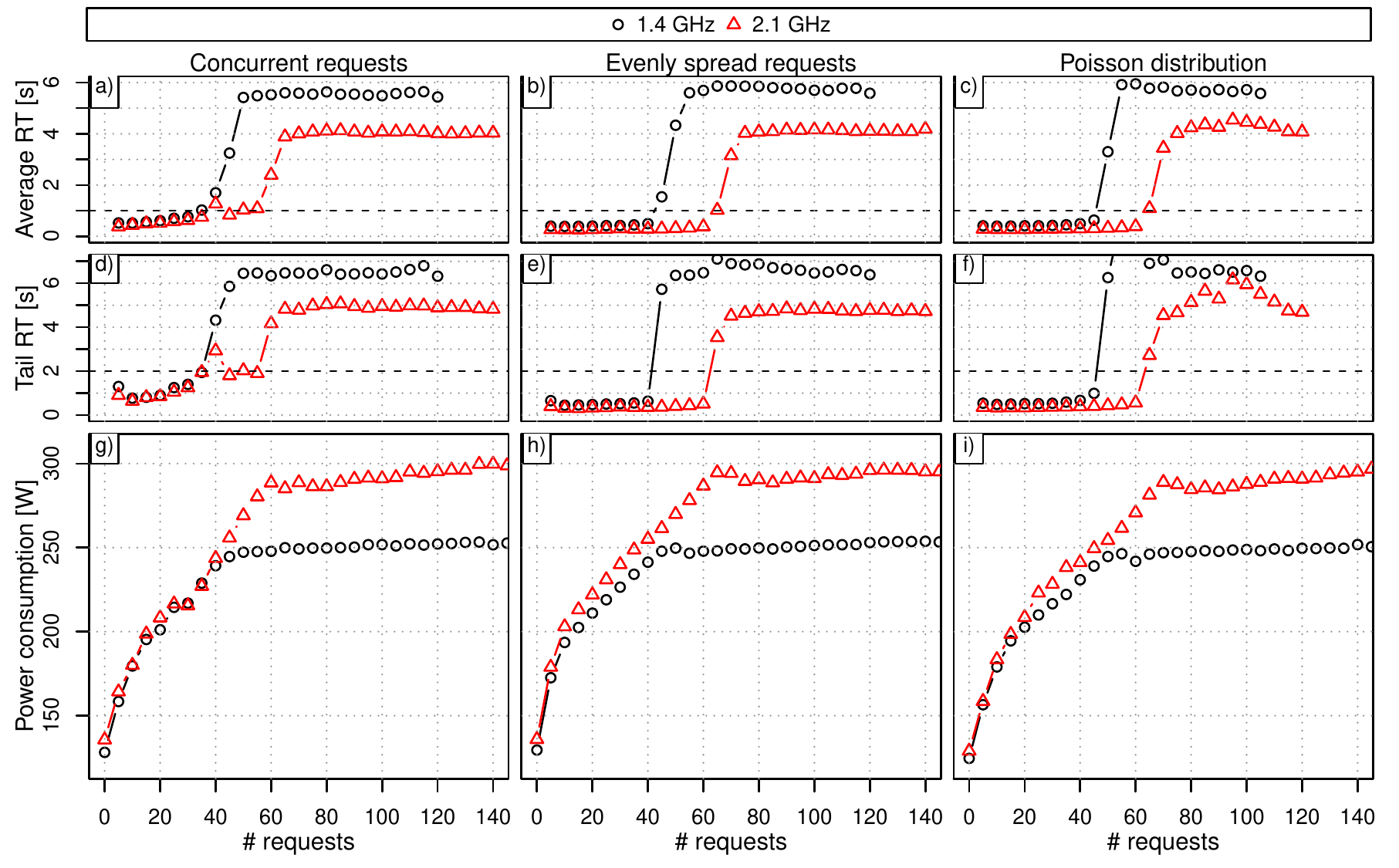}
    \caption{Influence of DVFS on the response time (RT) and power consumption for AMD servers.
    Dashed horizontal lines depict the observed thresholds of system saturation.
    The impact of DVFS on an underloaded server (up to 40 requests) depends on the request arrival pattern.
    The average and tail response time of an overloaded application stabilises because of the limit on not processed requests (see Section~\ref{sec:workload-generator}).}
    \label{fig:dvfs-memcached}
\end{figure*}

Figure~\ref{fig:dvfs-memcached} and Figure~\ref{fig:dvfs-intel} show the influence of workload intensity (x-axis), CPU frequency (data series), and request arrival pattern (each column) on the average response time (the first row), tail response time (the seconds row), maximal throughput (the intersect of data series and the dashed horizontal line), and the power consumption (the third row) for AMD and Intel servers.

\subsubsection{Influence of workload intensity}
Let us first consider the case of a physical server with AMD processor and a workload with evenly spread requests.
DVFS has an influence on the performance of the application, changing the size of workload that can be handled in underloaded state from 40 requests for 1.4~GHz to 60 requests for 2.1~GHz (Figure~\ref{fig:dvfs-memcached}b,e).
The difference in power consumption, shown in Figure~\ref{fig:dvfs-memcached}h, although statistically significant (see Table~\ref{tbl:dvfs-evenlydw-pc-t-test}) is much lower for an underloaded application (around 10~W) compared to the difference when the application is overloaded (around 40~W).
To compare power consumptions of two different settings we perform a Welch Two Sample t-test~\cite{welch1947generalization} that determines if the two populations have equal means.
We do not use the standard Student's t-test since the populations have unequal variances.
The difference in power consumption is considered significant when the $p$-value is lower than a significance level of 0.05.
\remark{when stating absolute power reductions in Watts, also include the relative reductions in percent.}

\begin{table}[t]
\centering
\caption{Welch Two Sample t-tests for DVFS (AMD, evenly spread requests)} \label{tbl:dvfs-evenlydw-pc-t-test}
\begin{tabular}{rllr}
\hline
\multirow{2}{*}{Requests}   & \multicolumn{2}{l}{Power consumption: $\mu$ ($\sigma$) [W]} & \multirow{2}{*}{$p$-value} \\ \cline{2-3}
                            & 1.4~GHz    		& 2.1~GHz        &                   \\ \hline
 5                          & 172.6 (25.9)      & 178.9 (28.0)          & 0.169             \\
10                          & 193.6 (4.4)       & 203.0 (4.1)           & \textless0.001    \\
15                          & 202.5 (4.4)       & 213.1 (4.4)           & \textless0.001    \\
\ldots                      & \ldots            & \ldots                & \ldots            \\
100                         & 251.2 (4.0)       & 291.2 (7.5)           & \textless0.001
\end{tabular}
\end{table}

\remark{I would cut Table I and say instead, that we have performed t-tests and have seen that in the text. I am also not sure why do we need a t-test over here. You might also want to look at such a test with the relative change (i.e., after subtracting the power consumption at idle), but I am not sure how important is that as it feels a bit artificial to me}

\textbf{Decreasing the CPU frequency of a server running an underloaded application reduces the power consumption only by approximately 10~W (\boldmath$\sim 5\%$) because lower frequency increases the time needed to process each request, and therefore the CPU spends more time in an active state}, as explained below.
\remark{This is confusing, does this mean that overloaded applications drop more requests (and can thus retain their power saving level)? What is the processing throughput relationship to all of this? After reading the next couple of sentences I get it, but the first parts are confusing, suggest to rephrase this to put the relationship first rather than last (in the paragraph). Also include something about what happens when the load cap is hit (even if it's just "requests get dropped").}
We use \emph{cpupower monitor} to measure processor idle statistic.
Table~\ref{tbl:active-and-idle-states-amd-evenly-spread} shows, for each frequency, the average ($\mu$) and the standard deviation ($\sigma$) of the time spent in active and idle CPU states calculated over all CPU cores assigned to the virtual machine.
The time CPU cores spend in an idle state, when hosting the underloaded MediaWiki with Memcached (5--45 requests), depends on the CPU frequency.
When operating on a lower CPU frequency, the application needs more time to serve each request, therefore CPU spends less time in the idle state.
However, when a CPU operates on a higher CPU frequency, requests are served in a shorter time, so the CPU can enter an idle state for a longer time till next requests arrive to the system, and therefore save more energy.

\begin{table}[t]
\centering
\caption{Time spent in the CPU active and idle states (AMD, evenly spread requests 5--45)} \label{tbl:active-and-idle-states-amd-evenly-spread}
\begin{tabular}{rrr}
\hline
\multirow{2}{*}{Frequency}   & \multicolumn{2}{c}{Time spent: $\mu$ ($\sigma$) [\%]} \\ \cline{2-3}
                & Active         & Idle          \\ \hline
1.4~GHz  & 57.2 (11.3)    & 42.8 (11.3)    \\
2.1~GHz  & 37.6 (12.5)    & 62.4 (12.5)       
\end{tabular}
\end{table}

When the application becomes overloaded, e.g., the workload higher than 40 requests for 1.4~GHz or 60 requests for 2.1~GHz at the AMD server, the response time increases rapidly and stabilises (Figure~\ref{fig:dvfs-memcached}b,e).
The average and tail response time of an overloaded application stabilises because of the limit on the number of not processed requests (see Section~\ref{sec:workload-generator}).
When there is no limit on the number of not processed requests, the response time would keep increasing rapidly till the system breaks (no successful responses).
The power consumption stabilizes because the server becomes saturated.

\begin{figure*}[t]
    \centering
    \includegraphics[width=\textwidth, trim={0 0 0 0},clip]{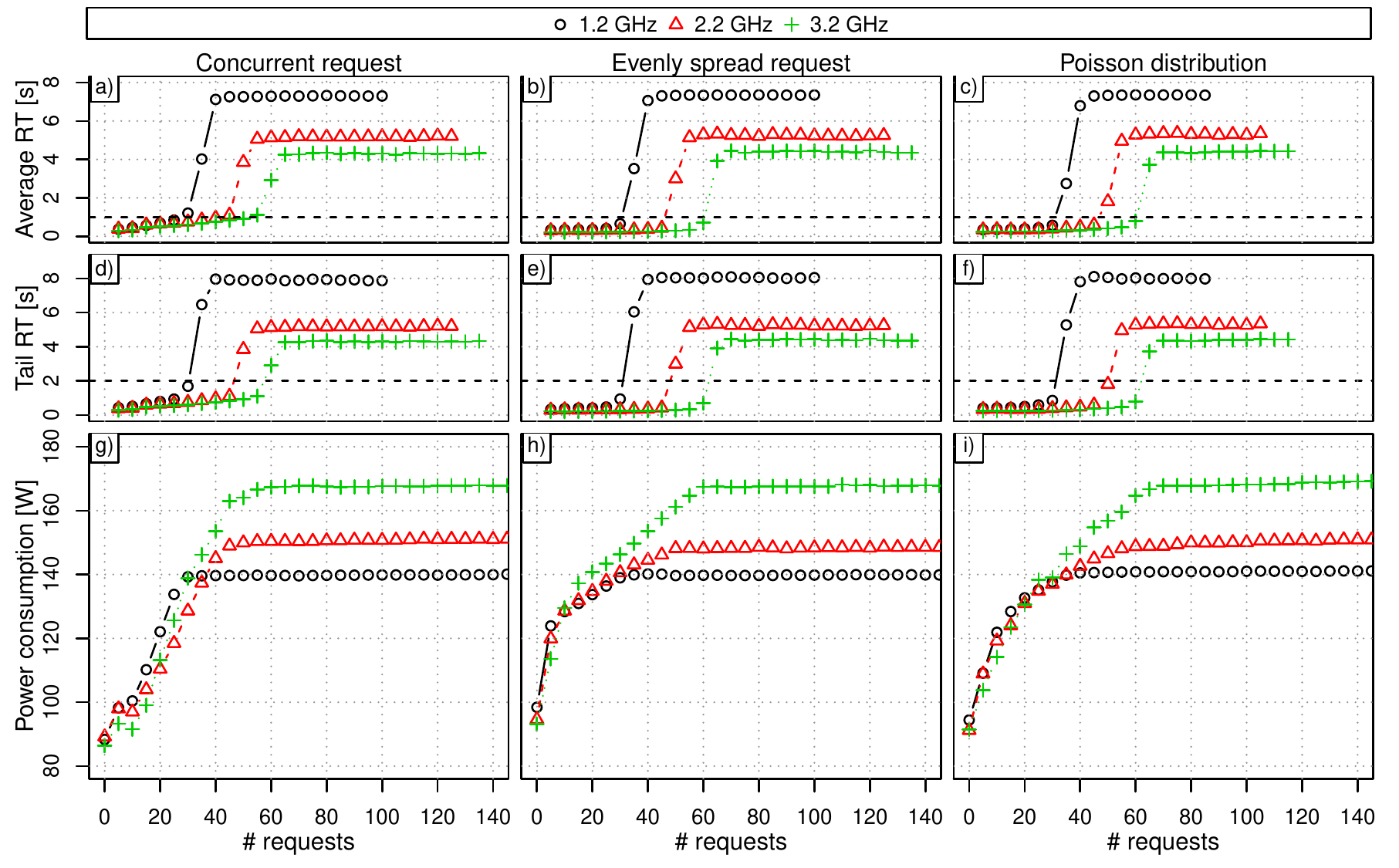}
    \caption{Influence of DVFS (Intel). For concurrent requests reducing CPU frequency causes and increase in the power consumption of an underloaded server.}\label{fig:dvfs-intel}
\end{figure*}

Next, we repeat the same experiment on the Intel server.
Since the drivers for DVFS of Intel processor allow us only to set the maximum allowed CPU frequency, the frequency may change between the minimum (1.2~GHz) and the one being set ($1.2$, $2.2$, or 3.2~GHz).
Figure~\ref{fig:dvfs-intel}a,d shows that the application behaves in a very similar manner (a lower number of requests overloads the application).
Also the difference in the power consumption between various CPU frequencies depends on the workload intensity (Figure~\ref{fig:dvfs-intel}g).
Moreover, one can observe that operating at the lowest CPU frequency (1.2~GHz) causes a significant performance degradation when the application is overloaded, increase in the response time of approximately 3~seconds while saving about 10~W.
On the other hand, using turbo frequency (3.2~GHz) reduces the response time about 1~second at the cost of approximately 20~W.

\remark{again, also include relative metrics (3s = 10\%? 100\%? 500\%?). Suggest to do this throughout the paper.}

Intel processors have multiple idle states with different power consumption and wake-up time.
Table~\ref{tbl:active-and-idle-states-intel-evenly-spread} shows how much time the CPU operating at 2.2~GHz spends in each state when the application starts to get overloaded (30--50 requests).
While the workload increases, the time that is spent in the idle states decreases.
Apart from that, also the distribution of time spent in different idle states changes.
While on the low workload (30 requests) the deepest idle state (C6-H) that reduces the power consumption the most, constitutes 97.15\% of the total time that CPU is idle.
When the application becomes overloaded the shallow idle states (e.g., C1E-), during which the power savings are smaller, start to occur more often.

\begin{table}[t]
\centering
\caption{Time spent in the CPU active and idle states (Intel, 2.2~GHz, evenly spread requests 5--30)} \label{tbl:active-and-idle-states-intel-evenly-spread}
\begin{tabular}{rrrrrrr}
\hline
\multirow{2}{*}{\# requests}   & \multicolumn{2}{c}{Average time spent [\%]} & \multicolumn{4}{c}{Percentage of idle [\%]} \\ \cline{2-7}
            & Active        & Idle (total)  & C1-H  & C1E-  & C3-H  & C6-H  \\ \hline
30          & 65.66         & 34.34         & 0.11  & 0.61  & 2.14  & 97.15 \\
35          & 76.59         & 23.42         & 0.09  & 1.72  & 4.67  & 93.52 \\
40          & 85.16         & 14.84         & 0.12  & 1.61  & 5.90  & 92.37 \\
45          & 93.53         & 6.48          & 0.10  & 3.19  & 7.72  & 88.98 \\
50          & 99.97         & 0.03          & 0.00  & 25.00 & 0.00  & 75.00 
\end{tabular}
\end{table}

\subsubsection{Influence of request arrival pattern}
\label{sec:DVFS-request-arrival-pattern}


Seeing the tradeoffs using a workload with evenly spread requests, we now study if changing the request arrival pattern has an effect on the power-performance tradeoffs.

When the requests are generated concurrently, the power consumption of an underloaded physical server
(with a workload up to 40 concurrent requests for 1.4~GHz and 60 requests for 2.1~GHz) is not significantly affected by the CPU frequency (Figure~\ref{fig:dvfs-memcached}g).
However, lowering the CPU frequency limits the power consumption when the server is
saturated.

Even though, due to the workload characteristics (requests arriving simultaneously), processing can be done in an uninterrupted manner, the proportions of total time spent in the CPU active and idle states are not affected (Table~\ref{tbl:active-and-idle-states}).
\remark{TODO Trevor: why the power consumption is lower when hosting an underloaded application for concurrently generated requests?}

\begin{table}[t]
\centering
\caption{Time spent in the CPU active and idle states (AMD, concurrent  requests 5--45)} \label{tbl:active-and-idle-states}
\begin{tabular}{rrr}
\hline
\multirow{2}{*}{Frequency}   & \multicolumn{2}{c}{Time spent: $\mu$ ($\sigma$) [\%]} \\ \cline{2-3}
                & Active        & Idle          \\ \hline
1.4~GHz  & 57.2 (6.2)    & 42.8 (6.2)    \\
2.1~GHz  & 41.6 (3.0)    & 58.4 (3.0)       
\end{tabular}
\end{table}

Table~\ref{tbl:dvfs-pc-t-test} shows, for each workload level,
the power consumption of a physical server operating at 1.4~GHz and 2.1~GHz, as well as, a $p$-value for the Welch Two Sample t-test.
The result of the Welch Two Sample t-tests shows that the means of power consumptions are not significantly different when the workload is lower than 45 requests.
Therefore, we conclude that DVFS is not able to reduce the power consumption of an underloaded physical server when the requests are generated concurrently.
However, for an overloaded application (workload higher that 40 requests) a difference in the power consumption becomes significant, as shown in the last two rows of Table~\ref{tbl:dvfs-pc-t-test} and in Figure~\ref{fig:dvfs-memcached} (bottom).

\begin{table}[t]
\centering
\caption{Welch Two Sample t-tests for DVFS (AMD, concurrent requests)} \label{tbl:dvfs-pc-t-test}
\begin{tabular}{rllr}
\hline
\multirow{2}{*}{Requests}   & \multicolumn{2}{l}{Power consumption: $\mu$ ($\sigma$) [W]} & \multirow{2}{*}{$p$-value} \\ \cline{2-3}
                            & 1.4~GHz    		& 2.1~GHz        &       \\ \hline
 5                          & 158.4 (31.3)      & 164.2 (31.9)          & 0.283 \\
10                          & 179.6 (26.9)      & 180.1 (31.5)          & 0.922 \\
15                          & 195.4 (20.1)      & 198.7 (25.2)          & 0.394 \\
20                          & 201.1 (19.9)      & 208.3 (28.3)          & 0.086 \\
25                          & 214.5 (16.2)      & 216.6 (27.3)          & 0.585 \\
30                          & 216.9 (18.5)      & 215.5 (34.2)          & 0.760 \\
35                          & 228.9 (13.2)      & 227.1 (33.7)          & 0.675 \\
40                          & 239.2 (12.0)      & 243.7 (38.4)          & 0.363 \\
45                          & 244.6 (8.4)       & 255.8 (35.6)          & 0.013 \\
50                          & 247.3 (4.9)       & 269.1 (23.2)          & \textless0.001\\
\ldots                      & \ldots            & \ldots                & \ldots            \\
100                         & 251.8 (5.0)       & 291.1 (8.7)           & \textless0.001
\end{tabular}
\end{table}

When request arrival pattern follows the Poisson distribution, the power consumption is not significantly affected by the DVFS for the workloads up to 15 requests per second (Figure~\ref{fig:dvfs-memcached}i).

\textbf{The request arrival pattern has a significant impact on the power consumption of an underloaded application, since it influences the number and duration of intervals that CPU cores spend in an idle state.}
When the requests are spread evenly along each second of the workload, the server with Intel processor consumes approximately 20~W more for a workload of 5 and 10 requests per second, comparing to the concurrent requests (see Figure~\ref{fig:dvfs-intel}h and Figure~\ref{fig:dvfs-intel}g).
In an extreme case of concurrently generated requests and an underloaded application, decreasing the CPU frequency even causes an increase in power consumption.
However, when the workload intensity increases, the difference due to the arrival pattern becomes less important, and finally, when the application becomes overloaded it does not have any measurable impact.

\remark{I find the line of reasoning a bit confusing, would probably be better to try to structure it from the point of view "DFVS is useful when" rather than the indirect line of reasoning used (that may be more intuitive the 5th time thinking about it, but not from the start)}

\section{CPU pinning}
\label{sec:cpu-pinning}

CPU pinning enables one to define a set of CPU cores that a process (or a virtual machine) is allowed to execute on.
It is possible to tie a process to a single CPU core, or to any subset of CPU cores.
In combination with the ability of the modern CPUs to put their unused parts into an idle state, CPU pinning technique could possibly be used to reduce the power consumption of the physical servers.


\subsection{Power consumption}
\begin{figure*}[t]
    \centering
    \begin{subfigure}[b]{0.23\textwidth}
        \includegraphics[width=\textwidth, trim={0 0 0 45},clip]{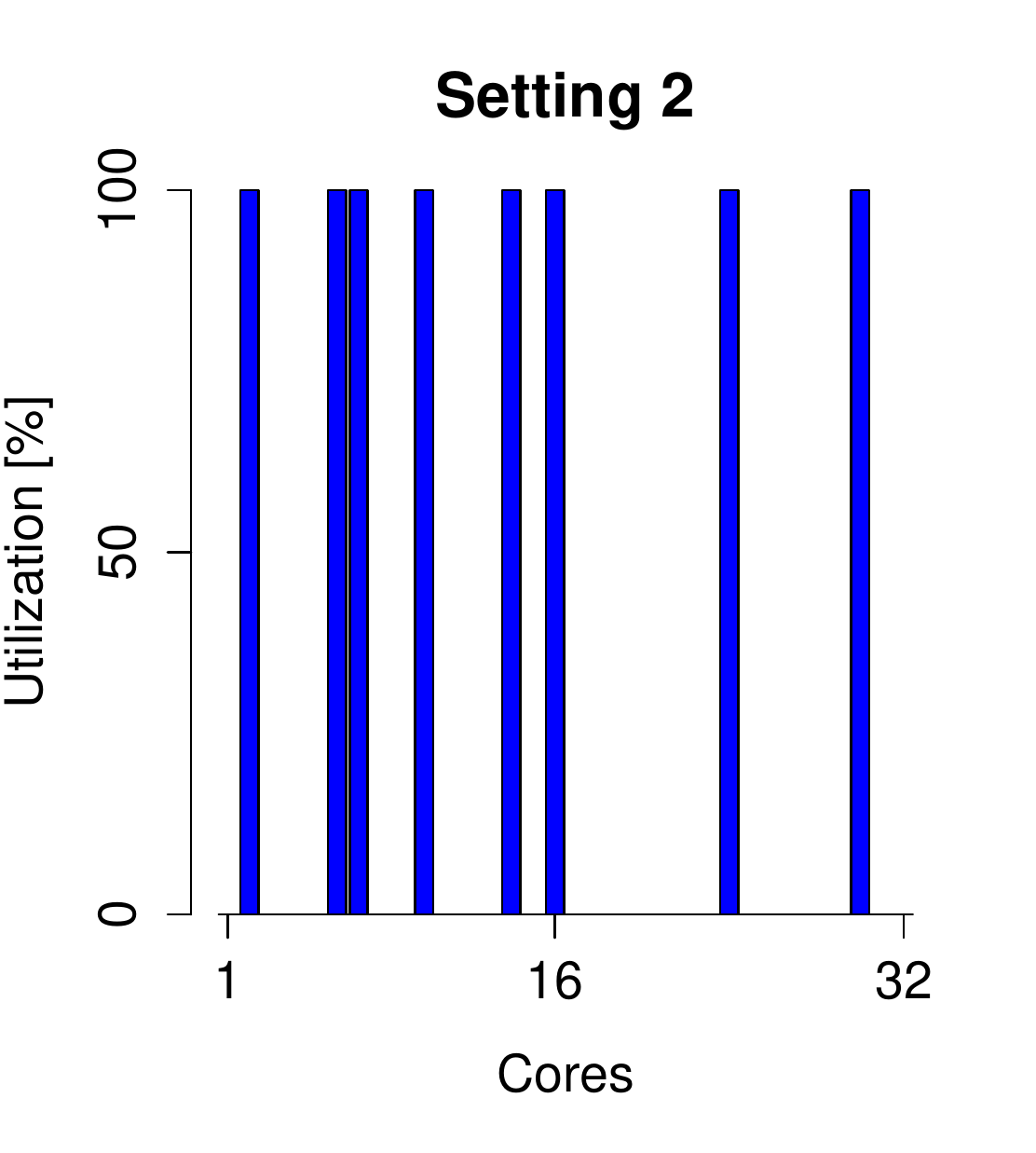}
        \caption{Setting 1}
        \label{fig:setting25_1}
    \end{subfigure}
    ~ 
    \begin{subfigure}[b]{0.23\textwidth}
        \includegraphics[width=\textwidth, trim={0 0 0 45},clip]{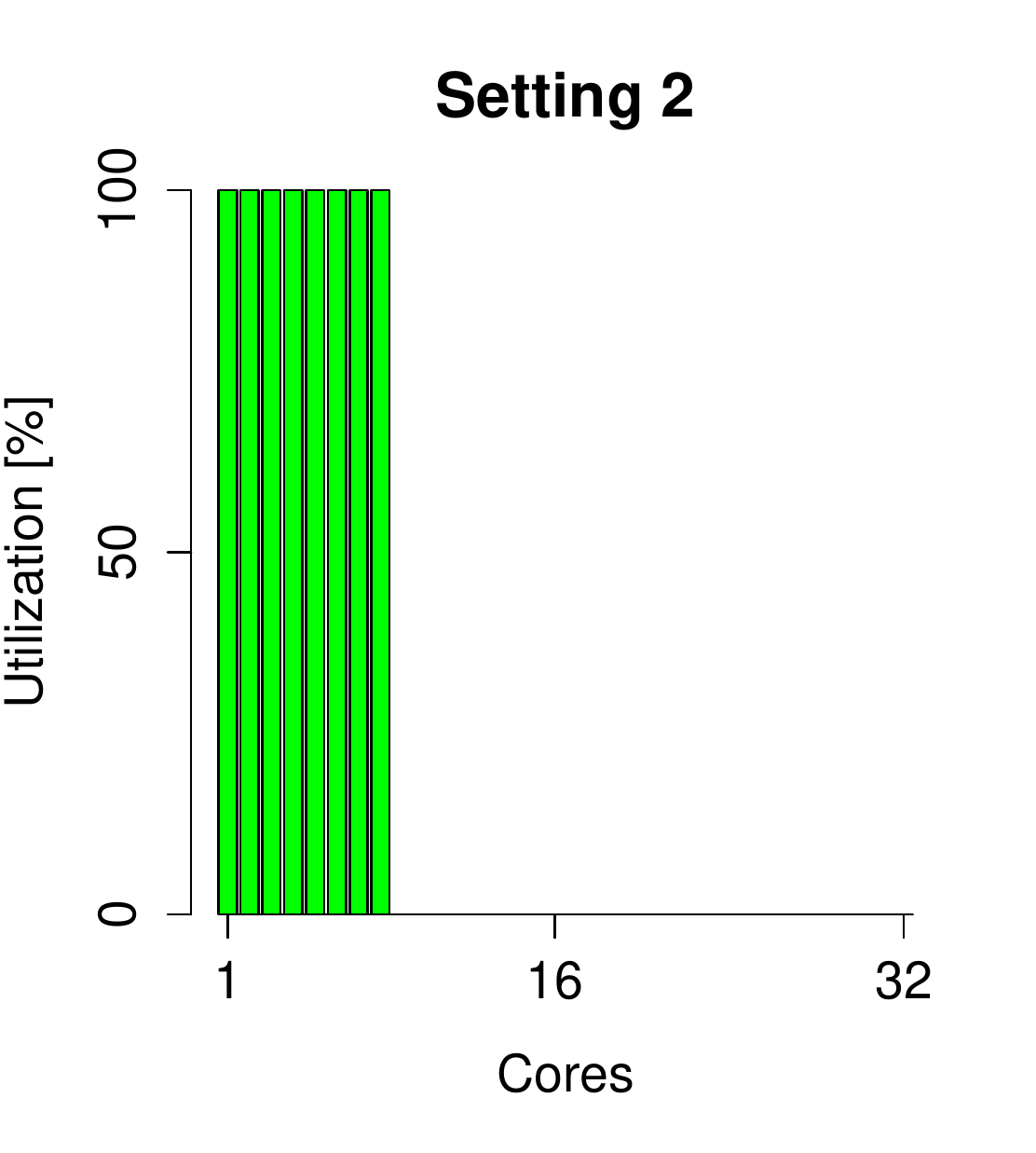}
        \caption{Setting 2}
        \label{fig:setting25_2}
    \end{subfigure}
    ~ 
    \begin{subfigure}[b]{0.23\textwidth}
        \includegraphics[width=\textwidth, trim={0 0 0 45},clip]{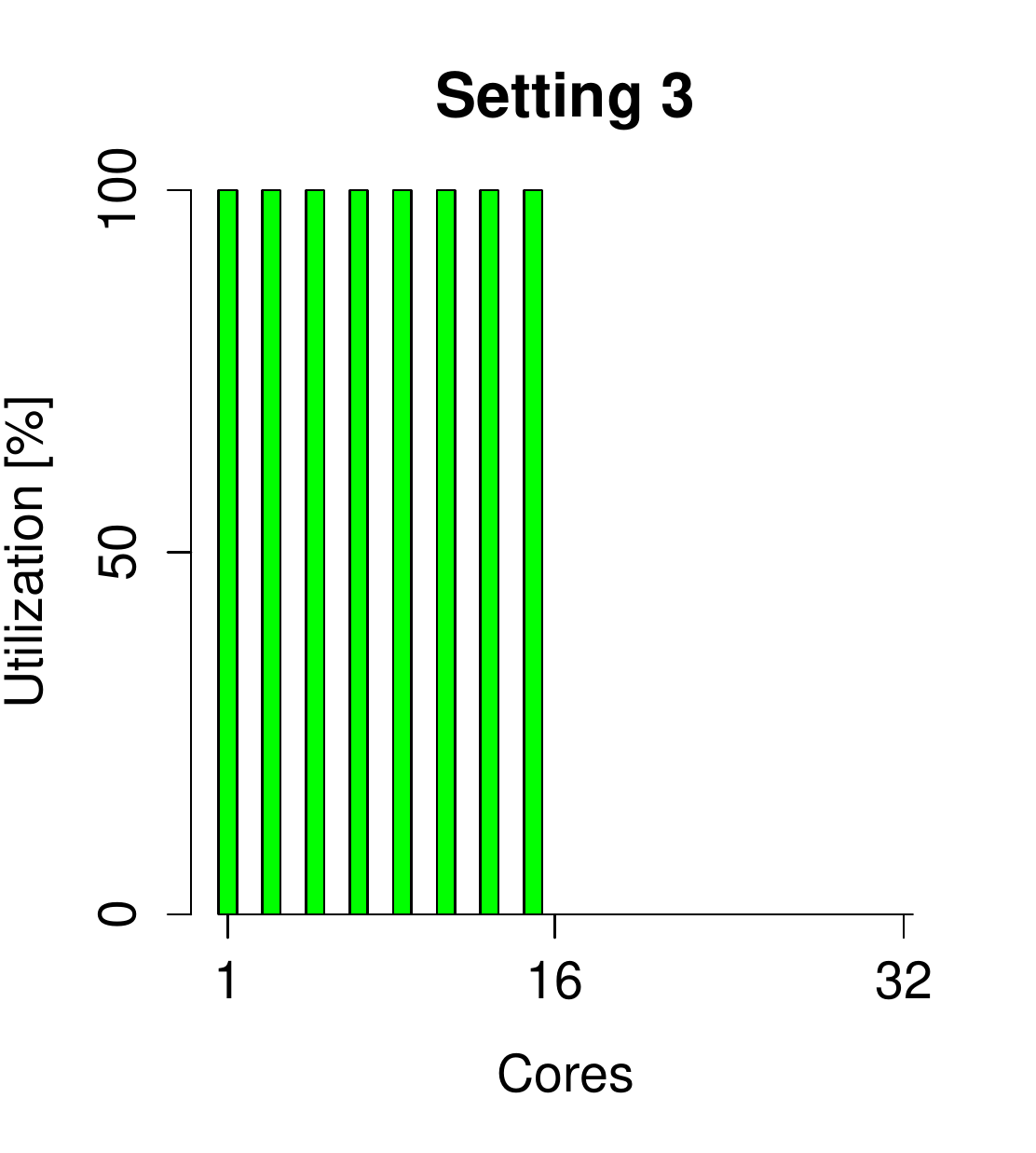}
        \caption{Setting 3}
        \label{fig:setting25_3}
    \end{subfigure}
    ~ 
    \begin{subfigure}[b]{0.23\textwidth}
        \includegraphics[width=\textwidth, trim={0 0 0 45},clip]{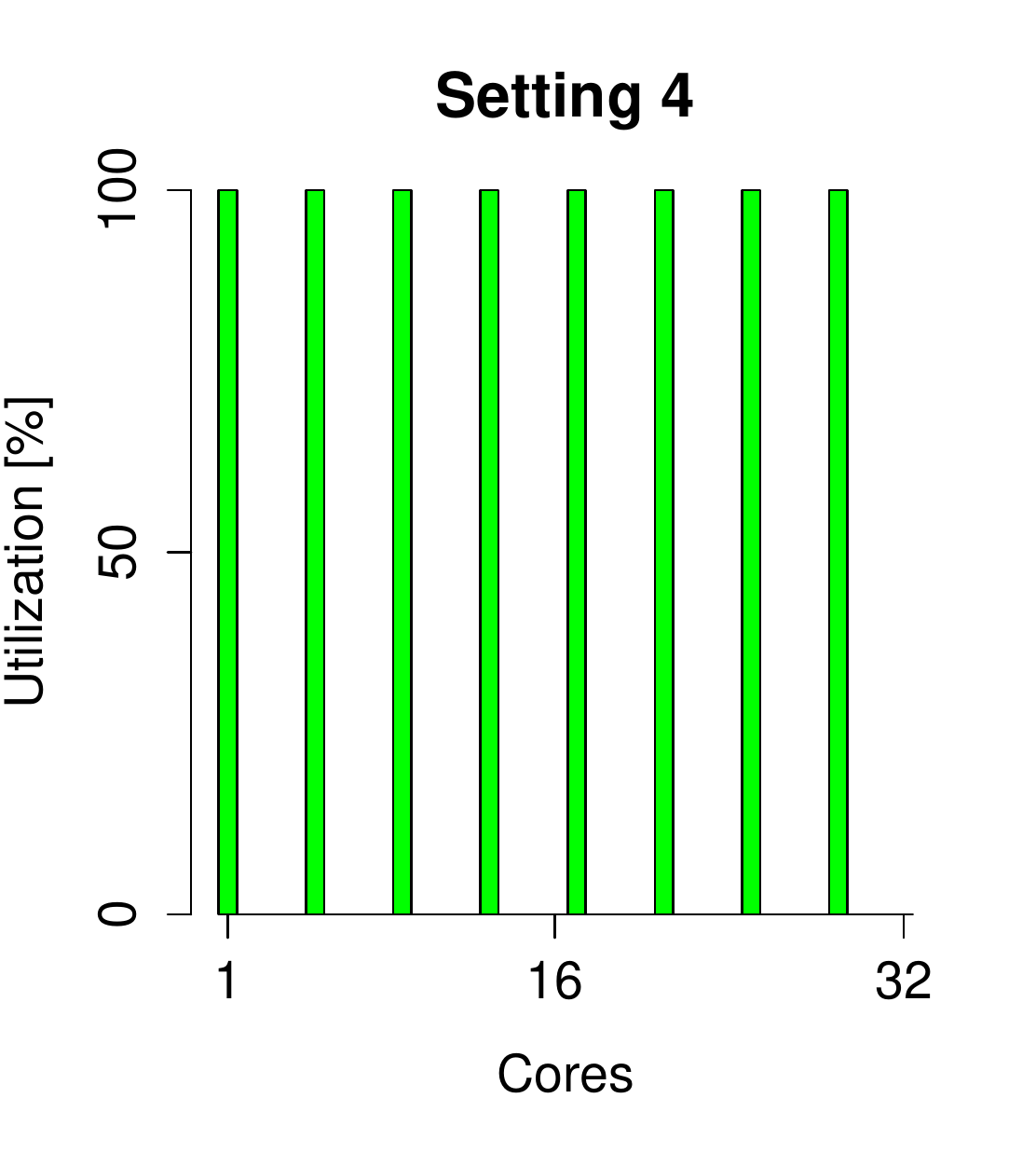}
        \caption{Setting 4}
        \label{fig:setting25_4}
    \end{subfigure} \\
        \begin{subfigure}[b]{0.23\textwidth}
        \includegraphics[width=\textwidth, trim={0 0 0 45},clip]{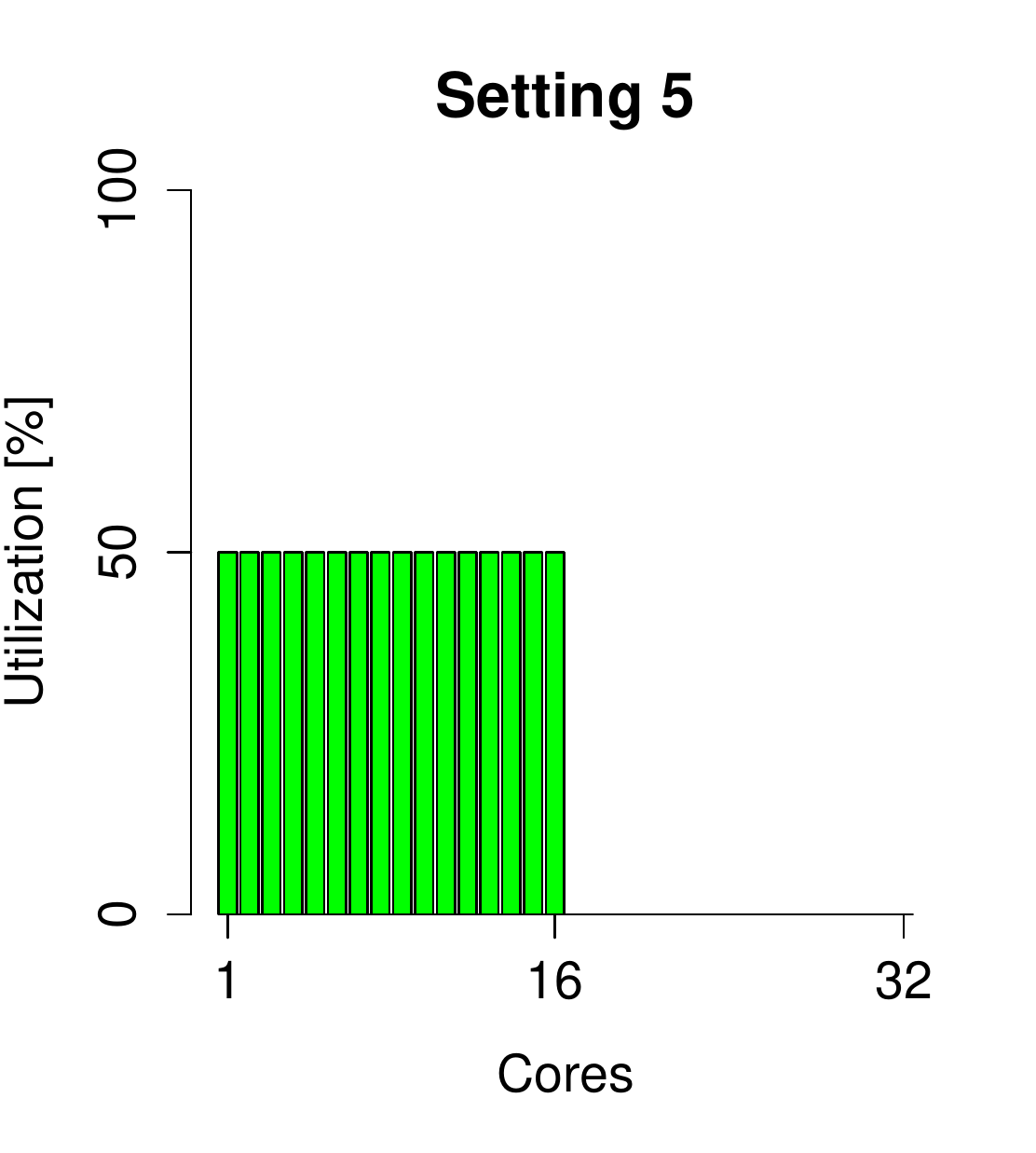}
        \caption{Setting 5}
        \label{fig:setting25_5}
    \end{subfigure}
    ~ 
    \begin{subfigure}[b]{0.23\textwidth}
        \includegraphics[width=\textwidth, trim={0 0 0 45},clip]{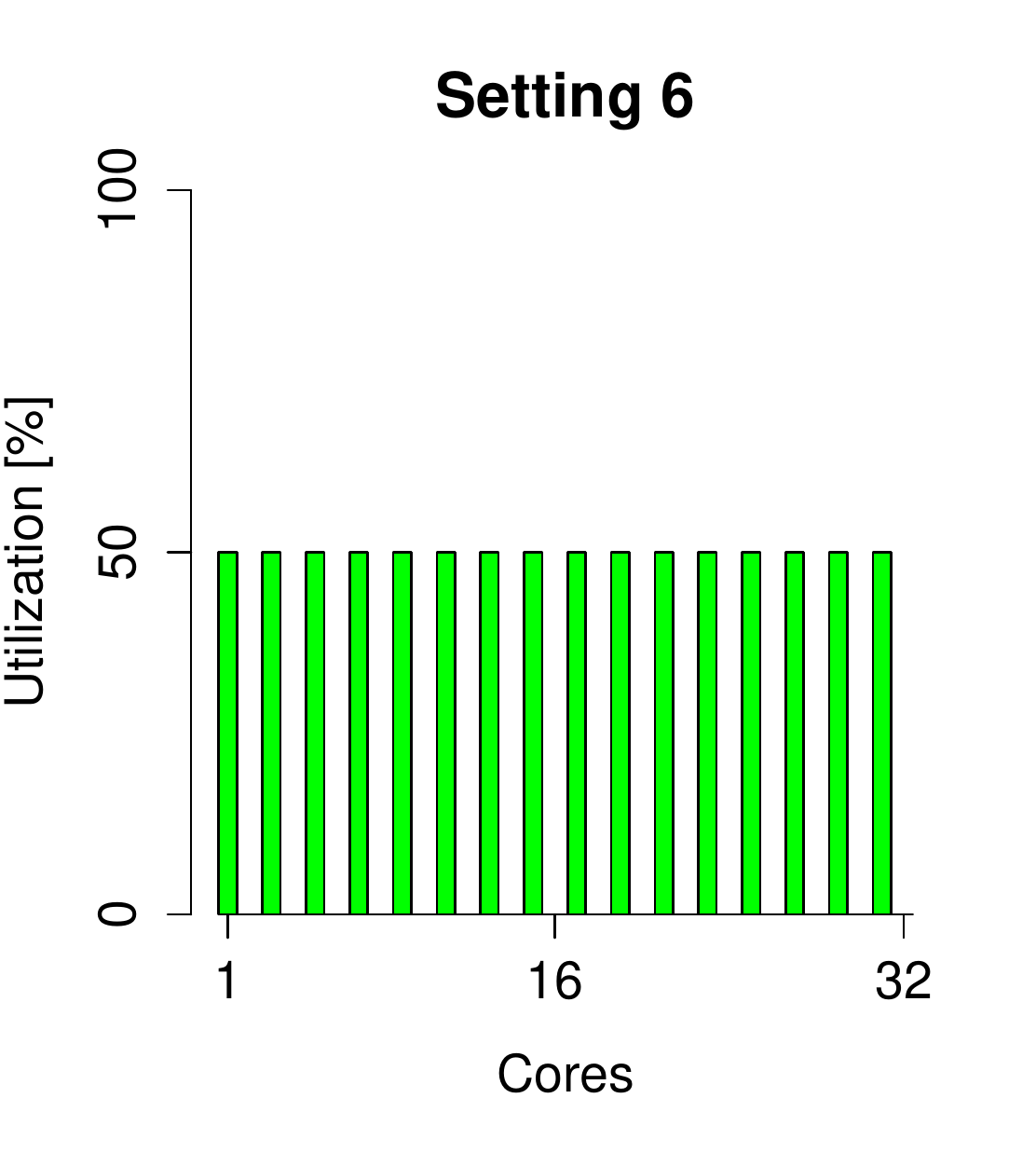}
        \caption{Setting 6}
        \label{fig:setting25_6}
    \end{subfigure}
    ~ 
    \begin{subfigure}[b]{0.23\textwidth}
        \includegraphics[width=\textwidth, trim={0 0 0 45},clip]{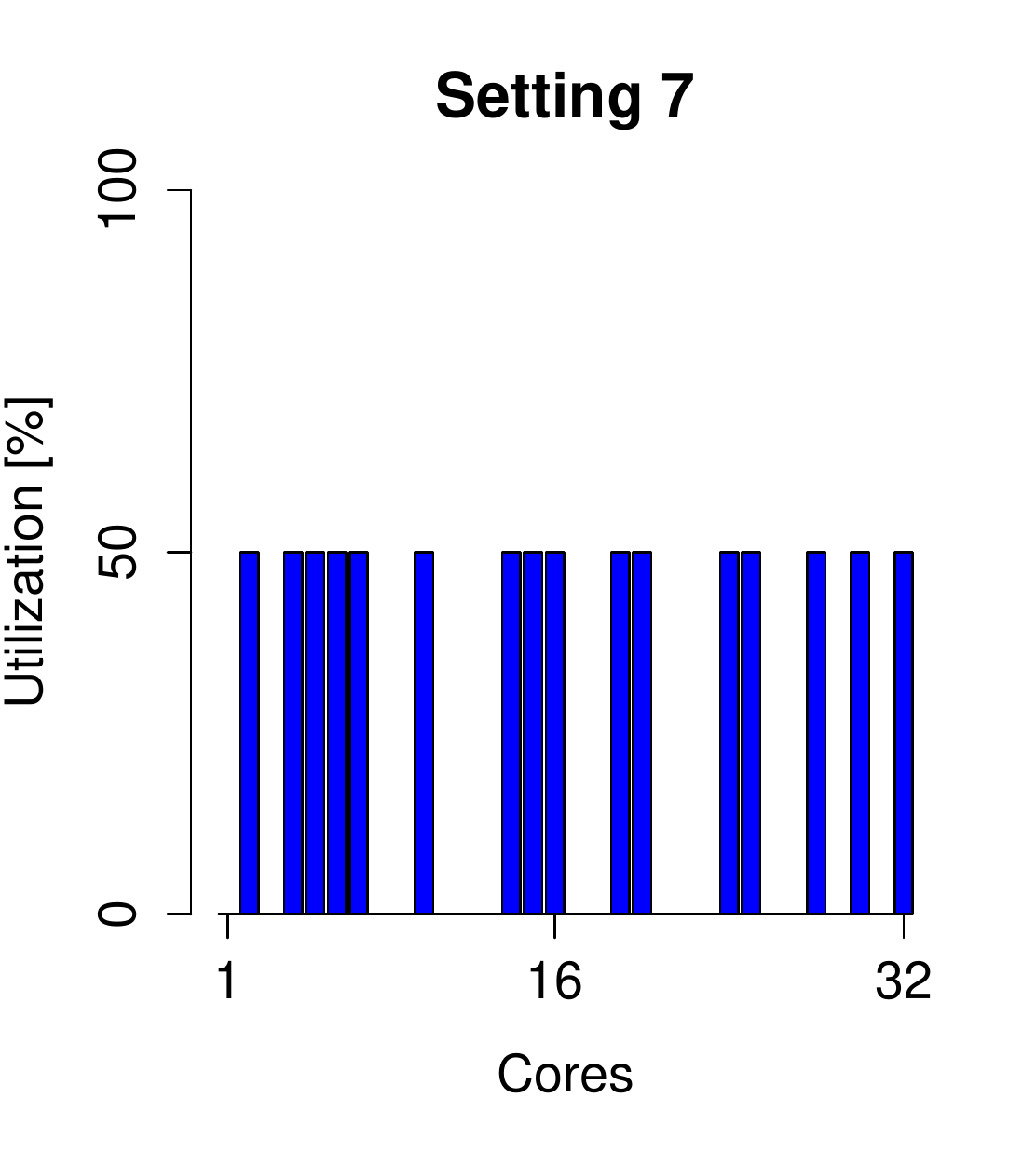}
        \caption{Setting 7}
        \label{fig:setting25_7}
    \end{subfigure}
    ~ 
    \begin{subfigure}[b]{0.23\textwidth}
        \includegraphics[width=\textwidth, trim={0 0 0 45},clip]{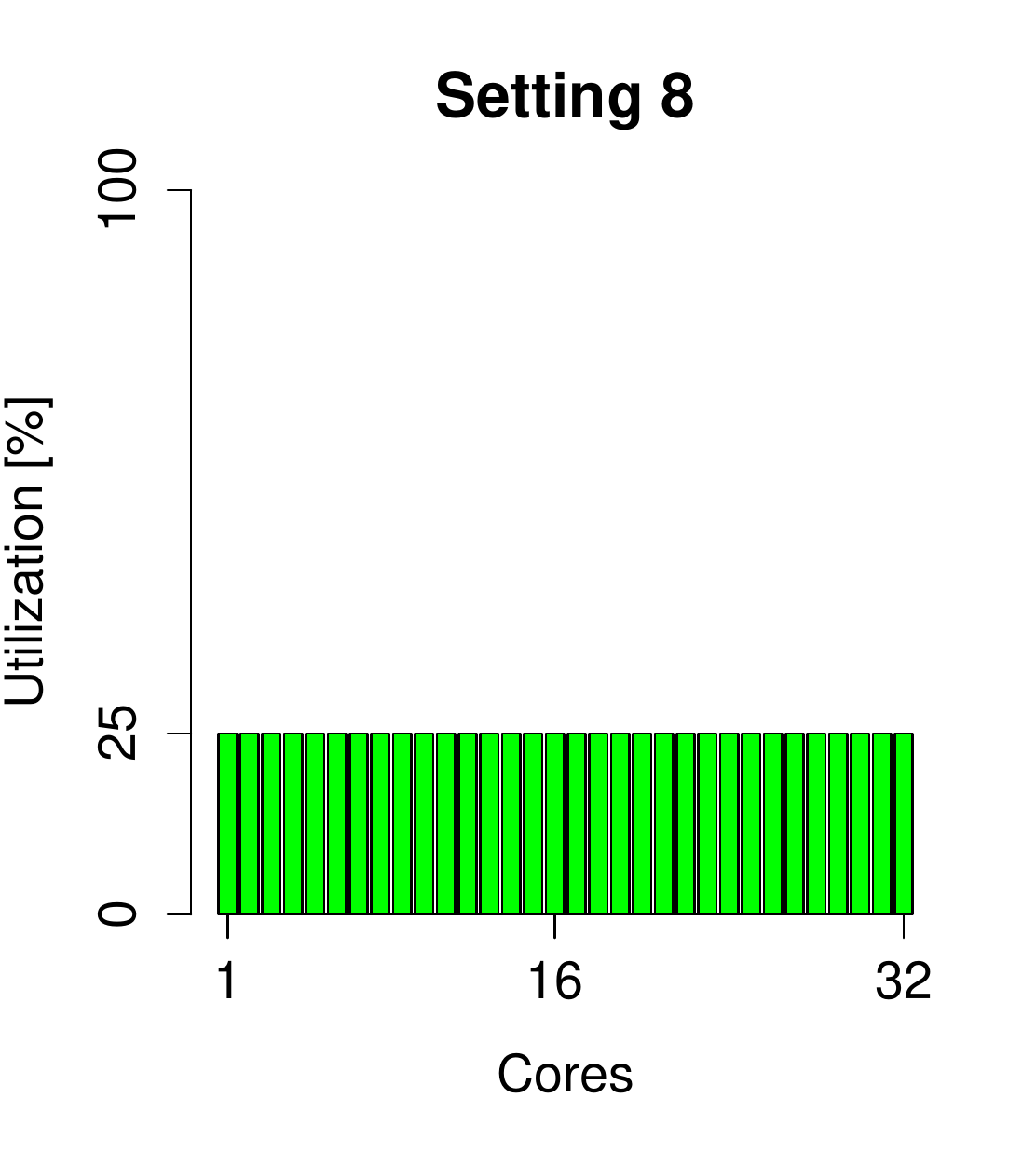}
        \caption{Setting 8}
        \label{fig:setting25_8}
    \end{subfigure}
    \caption{Experiment settings used for the evaluation of the influence of CPU cores consolidation using CPU pinning on a power consumption. Blue color means an assignment that changes during the experiment run, while green color means that the assignment is constant (processes are pinned to CPU cores).}\label{fig:experiment_settings25}
\end{figure*}

We start the analysis of how CPU pinning can be used to increase the energy efficiency by investigating how different arrangements of the same CPU utilization affect the power consumption of the AMD server (for the detailed description of the AMD processor architecture see Section~\ref{sec:testbed_servers}).
We compare eight settings with equal average CPU utilization (25\% of the total available resources) imposed using \emph{stress} and \emph{cpulimit} applications (running without virtualization).
Figure~\ref{fig:experiment_settings25} shows the CPU utilization levels on each core for all eight experiment settings.
Blue color means that processes are not pinned to particular cores and the mapping between processes and cores may change over the time of an experiment run.
Green color means that processes are pinned to cores and the mapping between processes and cores is fixed over the whole experiment run.

In the first four settings, eight cores are utilized at the level of 100\%.
In case of the first setting the cores are not pinned (Figure~\ref{fig:setting25_1}), while for settings 2--4 the core assignment is constant.
Utilized cores are packed on one CPU in Setting~2 (Figure~\ref{fig:setting25_2}), spread over one chip in Setting~3 (Figure~\ref{fig:setting25_3}), or over both chips in Setting~4 (Figure~\ref{fig:setting25_4}).
In the three next settings (5--7), sixteen cores are utilized at the level of 50\%.
Utilized cores are gathered on one chip in Setting~5 (Figure~\ref{fig:setting25_5}), spread across both chips in Setting~6 (Figure~\ref{fig:setting25_6}), or changing over time in Setting~7 (Figure~\ref{fig:setting25_7}). 
In the last setting (Figure~\ref{fig:setting25_8}) all cores are utilized at the level of 25\%.

\begin{figure}[t]
    \centering
    \includegraphics[width=0.6\textwidth, trim={0 5 20 30},clip]{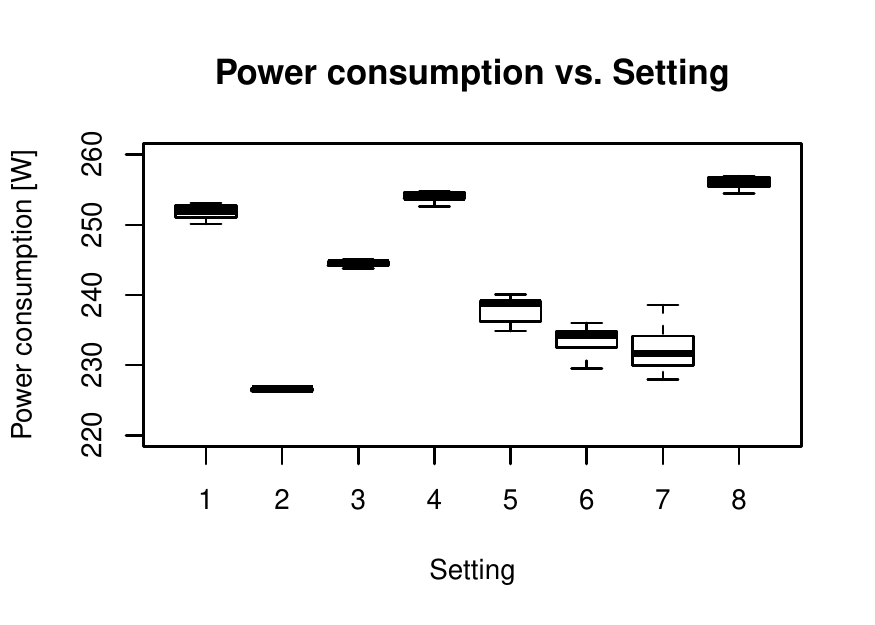}
    \caption{The relation between CPU pinning scheme and the power consumption.}
    \label{fig:settings25_box_plots}
\end{figure}

Figure~\ref{fig:settings25_box_plots} shows box plots with the distribution of average power consumption during experiment runs for each setting.
For settings 2--4, where eight cores are fully utilized, the power consumption is the lowest when all workload is processed on one CPU (Setting 2) and has a value of 227~W.
Spreading the workload across two CPUs on one chip (Setting 3) and across four CPUs on both chips (Setting 4) results in an increase of power consumption of 8\% and 12\% respectively.
It is caused by the heterogeneous CPU architecture, with some elements shared by groups of cores.
When even a single core from a group is active, it is required to provide power to the shared elements.
Therefore, spreading the workload increases the number of powered on shared elements and results in the increased total power consumption.
\remark{Idle power included? Maybe also talk about percentages when non-idle power is included.}

To check if CPU pinning can be used to reduce the average power consumption we conduct a one-way analysis of variance (one-way ANOVA) statistical test~\cite{fisher1925statistical} over the data presented in Figure~\ref{fig:settings25_box_plots}.
We do not use the t-test because we want to compare more than two settings.
One-way ANOVA works under an assumption that the data inside of each group is normally distributed.
We have verified that this is the case for our data using Shapiro-Wilk normality tests~\cite{shapiro1965analysis}.
Since there is a significant difference in variance between the settings (Bartlett's test $p$-value = $1.096 \cdot 10^{-7}$), we do not assume equal variances during the one-way ANOVA test and use the Welch correction for nonhomogeneity \cite{welch1951comparison}.
The result of the one-way ANOVA test ($p$-value = $2.2 \cdot 10^{-16}$) confirms that there is a significant difference between the averages.

This means that the way the processes are arranged on the CPU cores has an influence on the power consumption of a physical server.
The lowest power consumption that we have observed is for the configuration where the processes are pinned sequentially (Figure~\ref{fig:setting25_2}).
That strategy could be used to reduce the power consumption when then number of utilized cores decreases (e.g., after scale down in case of vertical scaling or scale in for horizontal scaling).
Therefore, we compare the relation between the number of utilized cores and the power consumption for two settings: processes can freely move across all CPU cores (default) and processes are pinned to the first $n$ cores.

\begin{figure}[t]
    \centering
    \includegraphics[width=0.6\textwidth, trim={0 5 20 30},clip]{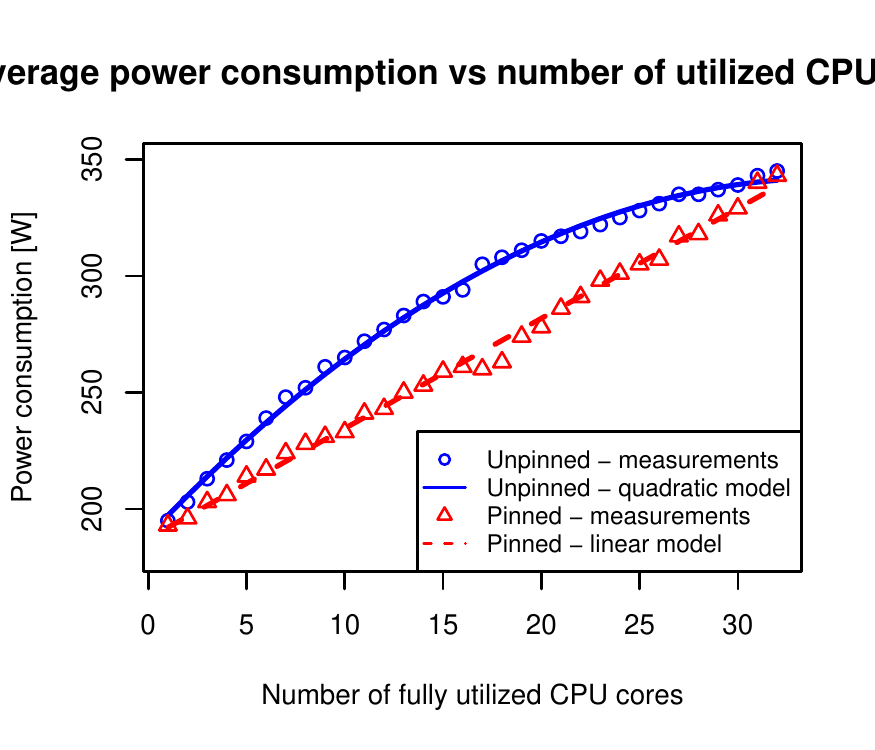}
    \caption{Pinning processes to succeeding cores makes a server dynamically energy proportional with respect to the number of fully utilized cores.}\label{fig:pinning-power-consumption}
\end{figure}

Figure~\ref{fig:pinning-power-consumption} shows the relation between the number of fully utilized CPU cores and the power consumption of a physical server.
\textbf{CPU pinning can be used to lower the power consumption for the mid utilization levels by enabling unused CPU cores to enter idle states, and even achieve the dynamic energy proportionality}~\cite{lo2014towards} -- a linear relation between the CPU utilization and the dynamic part of power consumption (not considering the idle power consumption).

When the processes are not pinned to a particular subset of CPU cores and they can move between CPU cores the relation between the number of fully utilized cores and the power consumption is not linear.
It can be modeled using a quadratic function $P=-0.13c^2+8.86c+188.53$, where $c$ is the number of fully utilized cores.
The regression model approximates the measurements very well, $R^2=0.998$.
From these quadratic model we can see that the cost of adding a CPU core decreases when the number of utilized cores grows.
Therefore, when admitting a new virtual machine, it is more power efficient to place it on an almost full physical server than on an almost idle one.

However, if the workload is pinned to the first $n$ CPU cores the relationship between the number of fully utilized cores and the power consumption becomes linear ($P=4.72c+187.38$, $R^2=0.995$).
In such a case, there is no difference in the power cost between placing a virtual machine on an almost idle physical server or on an almost full one.

\subsection{Power-performance tradeoff}
As we have shown in the previous section, the power consumption can be reduced by the consolidation of the virtual CPUs on a subset of physical CPU cores.
In this section, we evaluate how the virtual CPU consolidation affects the application performance.

We run experiments with a system consisting of a load balancer and 8 instances of the MediaWiki application (1 CPU core and 2~GB of RAM per virtual machine).
Virtual CPUs are pinned in the same way as we have pinned the processes in the previous experiment (Figure~\ref{fig:experiment_settings25}).
The system is exposed to a constant workload (32 requests per second), and the application performance as well as the power consumption of the physical server is monitored.

\begin{figure}[t]
    \centering
    
    \includegraphics[width=0.6\textwidth, trim={0 0 0 0},clip]{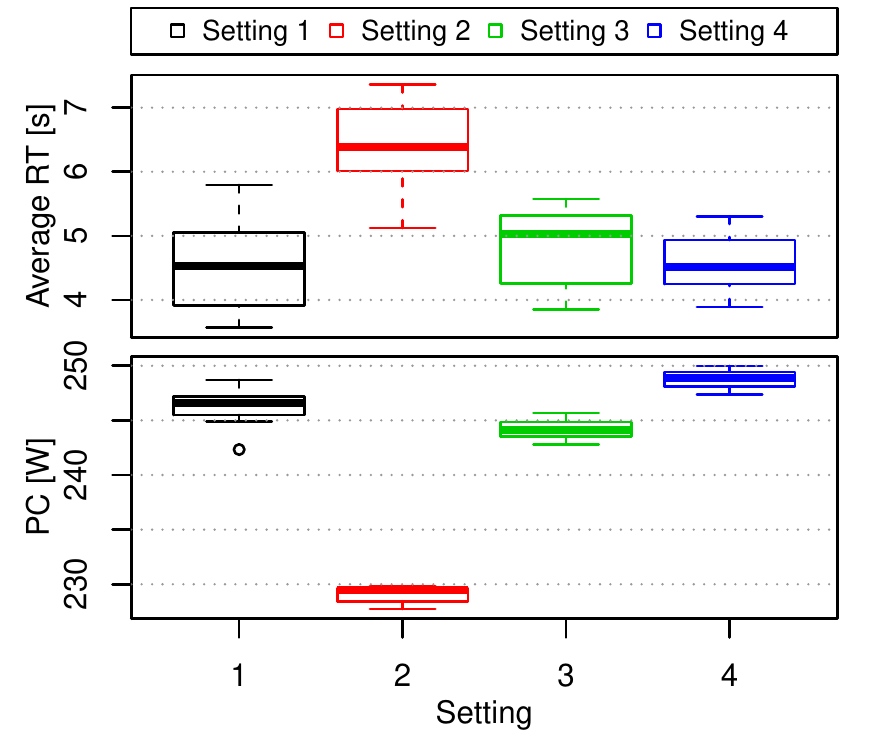}
    
    \caption{Influence of the CPU pinning schemes on the power consumption (PC) of a physical server and the application response time (RT) for a saturated system.}\label{fig:wiki_horizontal}
\end{figure}

Figure~\ref{fig:wiki_horizontal} shows that pinning of virtual machines' virtual cores to the physical CPU cores affects both the power consumption of physical server and the performance of the applications that are running inside the virtual machines.
Pinning all virtual machines to the first $n$ consecutive cores gives the lowest power consumption, but at the same time increases response time due to the contention of shared resources (e.g., memory bandwidth, last level cache).

\subsubsection{Influence of workload intensity}
\begin{figure}[t]
    \centering
    
    \includegraphics[width=0.6\textwidth, trim={0 0 0 0},clip]{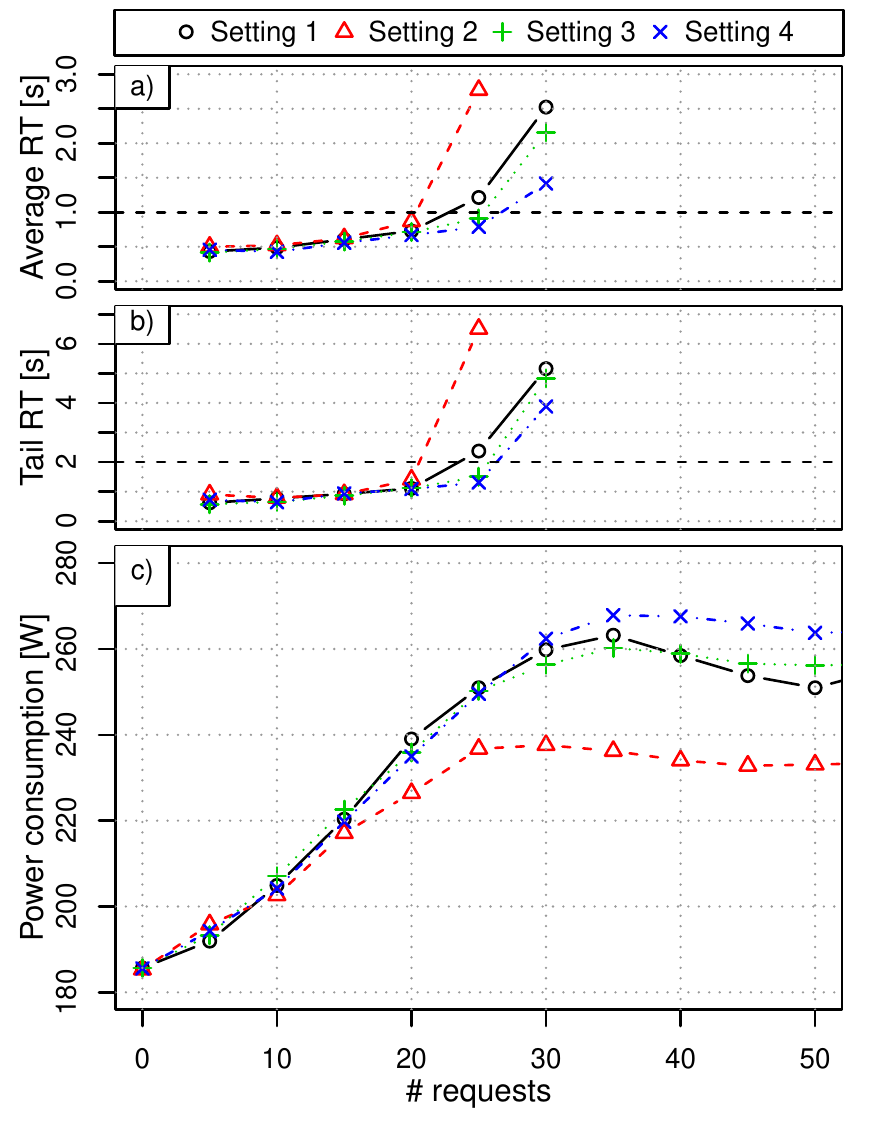}
    
    \caption{Influence of the CPU pinning on the power consumption of the AMD physical server and the application response time (RT) depending on the workload intensity.}
    \label{fig:memcached_cpu_pinning}
\end{figure}

Next, we run experiments in which MediaWiki is hosted in a single virtual machine with 8 physical CPU cores allocated. 
MediaWiki is exposed to a step workload ranging from 0 to 50 requests per second.
The CPU pinning schemes are changed between experiments similarly to the settings 1--4 shown in \mbox{Figure~\ref{fig:experiment_settings25}a--d}.
Figure~\ref{fig:memcached_cpu_pinning} shows the influence of different CPU pinning schemes on the performance of MediaWiki with Memcached and the power consumption of a physical server hosting the virtual machine.

To verify if the power consumption of a physical server can be reduced using virtual CPU consolidation (pinning virtual CPUs to the first $n$ physical CPU cores) we conduct a series of Welch Two Sample t-tests over the power consumption measurements presented in Figure~\ref{fig:memcached_cpu_pinning}.
Table~\ref{tbl:cpu-pinning-pc-t-test} shows, for each workload level, the average ($\mu$) and the standard deviation ($\sigma$) of power consumption for the configuration without CPU pinning (Setting~1) and the configuration when all virtual CPU cores are pinned to the first $n$ physical CPU cores (Setting 2), as well as, a $p$-value for the Welch Two Sample t-test.

\begin{table}[t]
\centering
\caption{Welch Two Sample t-tests for CPU pinning} \label{tbl:cpu-pinning-pc-t-test}
\begin{tabular}{rllr}
\hline
\multirow{2}{*}{Requests}   & \multicolumn{2}{l}{Power consumption: $\mu$ ($\sigma$) [W]} & \multirow{2}{*}{$p$-value} \\ \cline{2-3}
                            & Setting 1         & Setting 2             &       \\ \hline
 5                          & 191.9 (9.3)       & 195.8 (9.9)           & 0.125 \\
10                          & 204.9 (17.1)      & 202.6 (13.7)          & 0.567 \\
15                          & 220.4 (19.8)      & 217.1 (11.4)          & 0.440 \\
20                          & 239.0 (18.6)      & 226.4 (9.6)           & 0.003 \\
\ldots                      & \ldots            & \ldots                & \ldots            \\
50                          & 251.0 (2.4)       & 233.1 (0.8)           & \textless0.001
\end{tabular}
\end{table}

The results of the Welch Two Sample t-tests show that the consolidation of virtual CPUs through CPU pinning does not significantly change the power consumption for the underloaded application (5--15 requests).
However, when the workload approaches the level when the application becomes overloaded (20 requests) the difference in the power consumption between CPU pinning schemes becomes significant.

\subsubsection{Influence of request arrival pattern}

The influence of request arrival pattern is not as significant as in case of DVFS (Section~\ref{sec:DVFS-request-arrival-pattern}).
When the requests are evenly spread, the application is able to handle slightly higher workloads and the response times are lower when the application becomes overloaded.
The power consumption in not affected by the request arrival pattern.

\subsubsection{Influence of shared CPU components}
To evaluate how 
the virtual CPU consolidation
affects the performance of two collocated virtual machines on the AMD server we pin one virtual machine to the first physical CPU core (core 0) and move the second virtual machine across all the other physical CPU cores (core 1--31).
When both virtual machines are pinned to the first two cores (cores 0 and 1) the average and tail response times are significantly higher than in the other cases.
The performance degradation in this configuration is most probably caused by the contention of shared resources: L1 and L2 instruction cache misses, last level caches, and DRAM bandwidth.
The results show that the CPU architecture should be taken into account when choosing a CPU pinning scheme to avoid a negative impact on the application performance.

\subsubsection{Influence of the number of utilized CPU cores}

\begin{figure}[t]
    \centering
    \begin{subfigure}[b]{0.5\textwidth}
        \includegraphics[width=\linewidth, trim={0 0 0 45},clip]{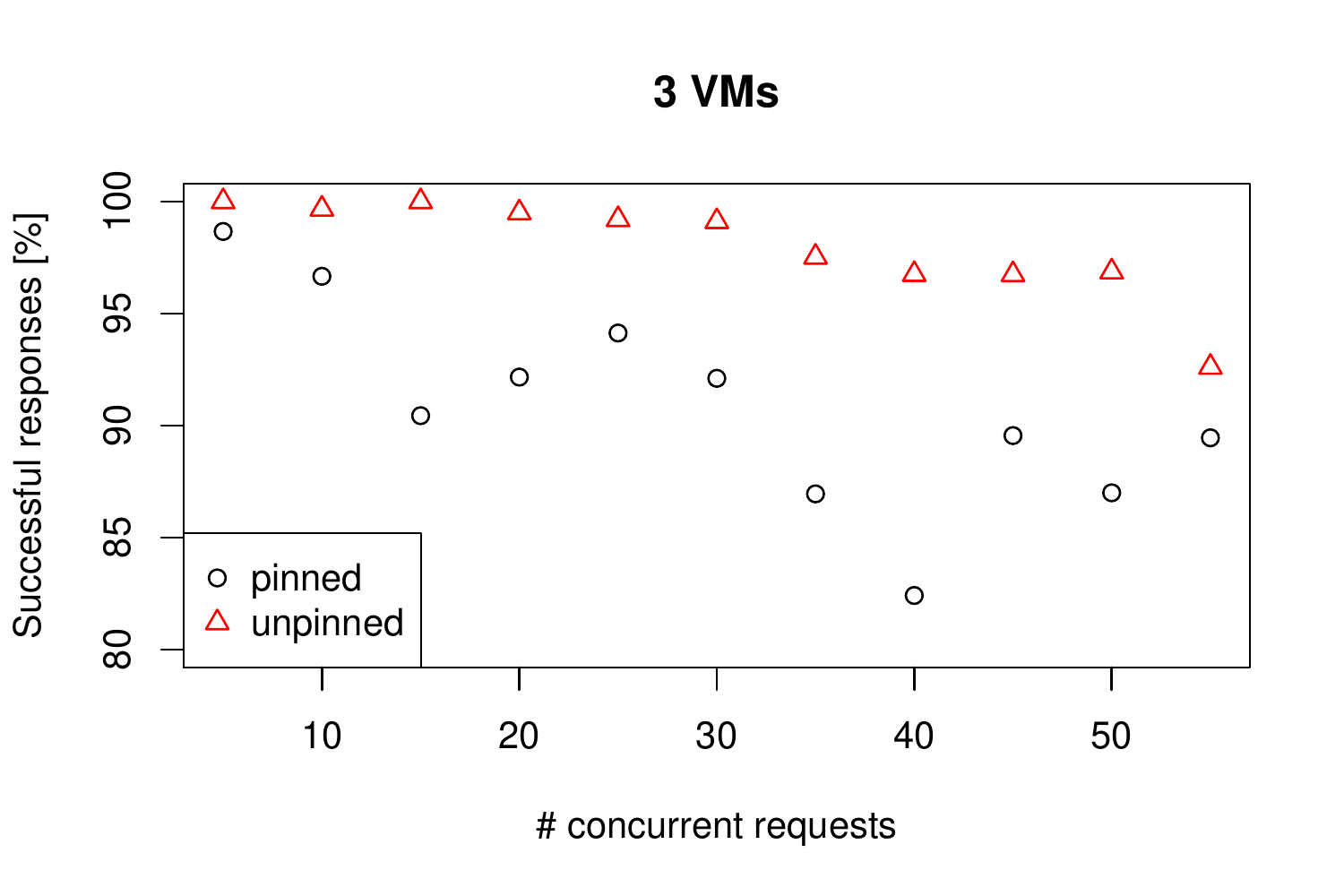}
        \caption{Success ratio for 3 virtual machines}
        \label{fig:wiki_pinning_horizontal_elasticity_3vms_sr}
    \end{subfigure}%
    \begin{subfigure}[b]{0.5\textwidth}
        \includegraphics[width=\linewidth, trim={0 0 0 45},clip]{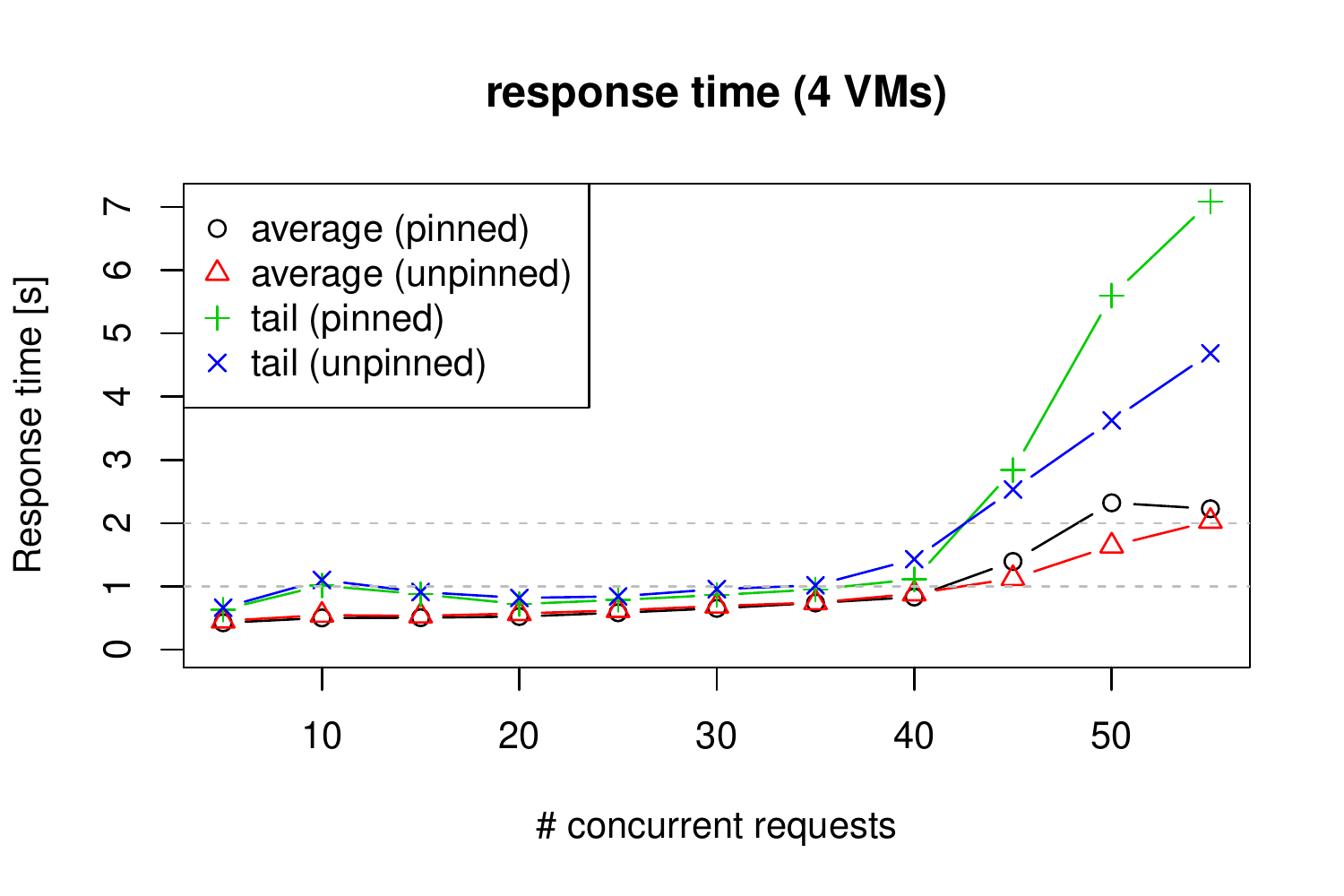}
        \caption{Response time for 4 virtual machines}
        \label{fig:wiki_pinning_horizontal_elasticity_4vms_rt}
    \end{subfigure}
    
    \begin{subfigure}[b]{0.5\textwidth}
        \includegraphics[width=\linewidth, trim={0 0 0 45},clip]{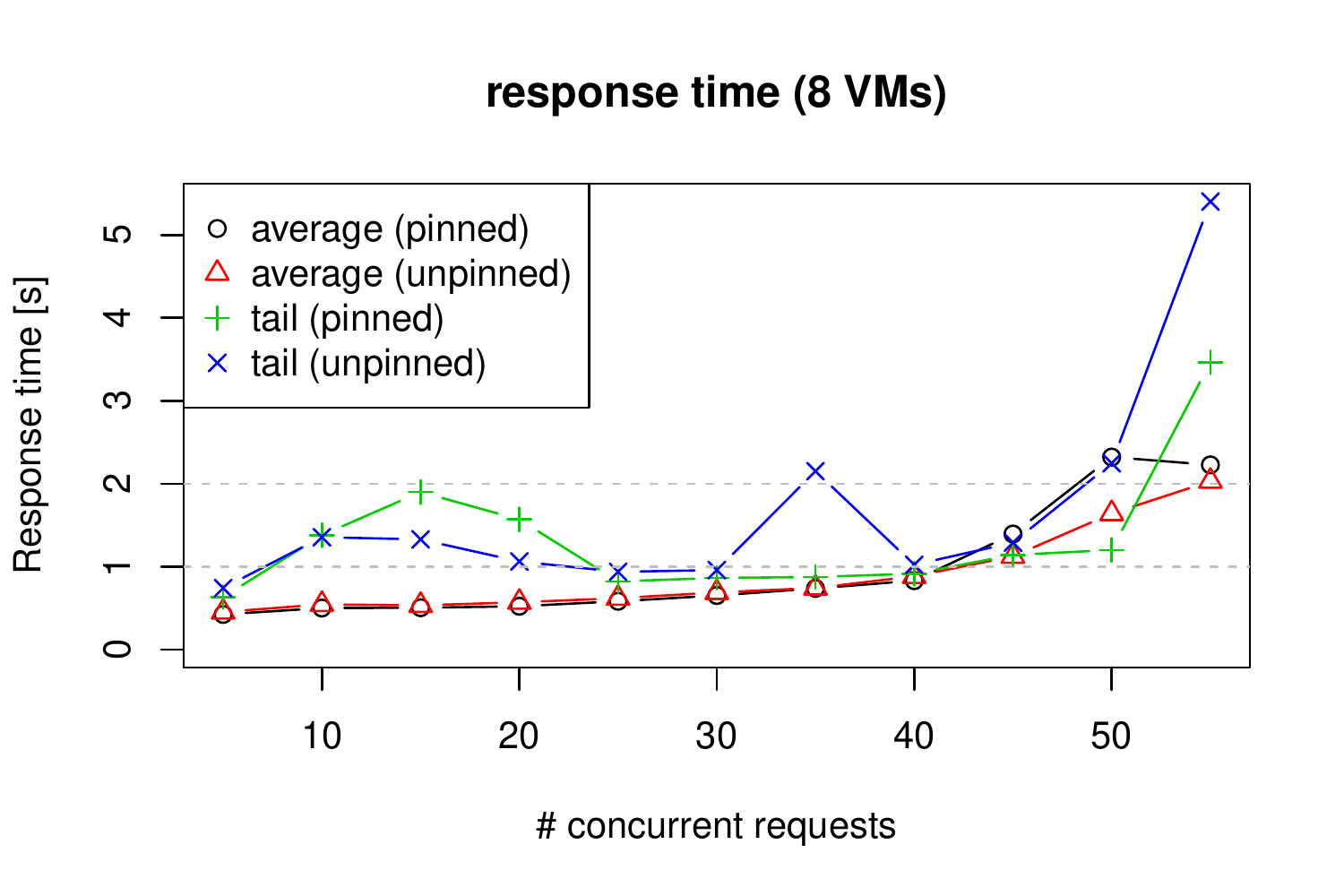}
        \caption{Response time for 8 virtual machines}
        \label{fig:wiki_pinning_horizontal_elasticity_8vms_rt}
    \end{subfigure}
    
    \caption{Influence of CPU pinning.}\label{fig:wiki_pinning_horizontal_elasticity_sr}
\end{figure}

We observe that virtual CPU consolidation affects the performance differently depending on the number of utilized physical CPU cores and application requirements.
We compare the influence of CPU pinning on the performance of a system consisting of a load balancer and multiple instances (1--8) of the MediaWiki application hosted in virtual machines with four CPU cores and 1.9 GB of RAM each.
Figure~\ref{fig:wiki_pinning_horizontal_elasticity_3vms_sr} shows a negative impact of the CPU pinning on a system with small processing capabilities (three virtual machines).
Pinning of the virtual CPUs to the first twelve physical CPU cores results in client not receiving responses to up 17\% of requests, due to 
internal server errors (HTTP response 500),
service unavailability (HTTP response 503),
or time-out on client side (10 seconds).
For a system with medium processing capabilities (four virtual machines) pinning the virtual CPUs to the first sixteen physical CPU cores increases the average and tail response times for a workload with more than 40 requests (see Figure~\ref{fig:wiki_pinning_horizontal_elasticity_4vms_rt}).
The performance degradation is caused by congesting the shared resources due to pinning virtual CPUs close to each other. 
\textbf{However, for a system with using all the CPU cores available on the physical server, CPU pinning improves the application response time by eliminating rotations of processes among cores that cause unnecessary cache repopulations} (Figure~\ref{fig:wiki_pinning_horizontal_elasticity_8vms_rt}).
In that case, CPU pinning does not reduce the number of allowed CPU cores, so it does not introduce congestion on the shared resources.
Moreover, CPU pinning eliminates the performance degradation due to migrations of virtual machines between different NUMA nodes.


\remark{Figure Influence of CPU pinning. -- I read this as, more VMs not useful at all?}

\section{Horizontal and Vertical Scaling}
\label{sec:scaling}

Changing the number of virtual machines hosting application instances (horizontal scaling) or the amount of resources allocated to a virtual machine (vertical scaling) can be
done
to adjust the system computational capabilities.
Since both horizontal and vertical scaling change the amount of physical resources used to serve the application workload, they have an indirect influence on the power consumption of servers hosting the virtual machines.

To investigate the power-performance tradeoffs of vertical and horizontal scaling, we perform a set of experiments using the MediaWiki application with Memcached.
We evaluate the costs and benefits of scaling the application vertically (with respect to the number of CPU cores and size of RAM allocated to a single virtual machine) and horizontally (with respect to the number of virtual machines).

\begin{figure*}[t]
    \centering
    \includegraphics[width=\textwidth]{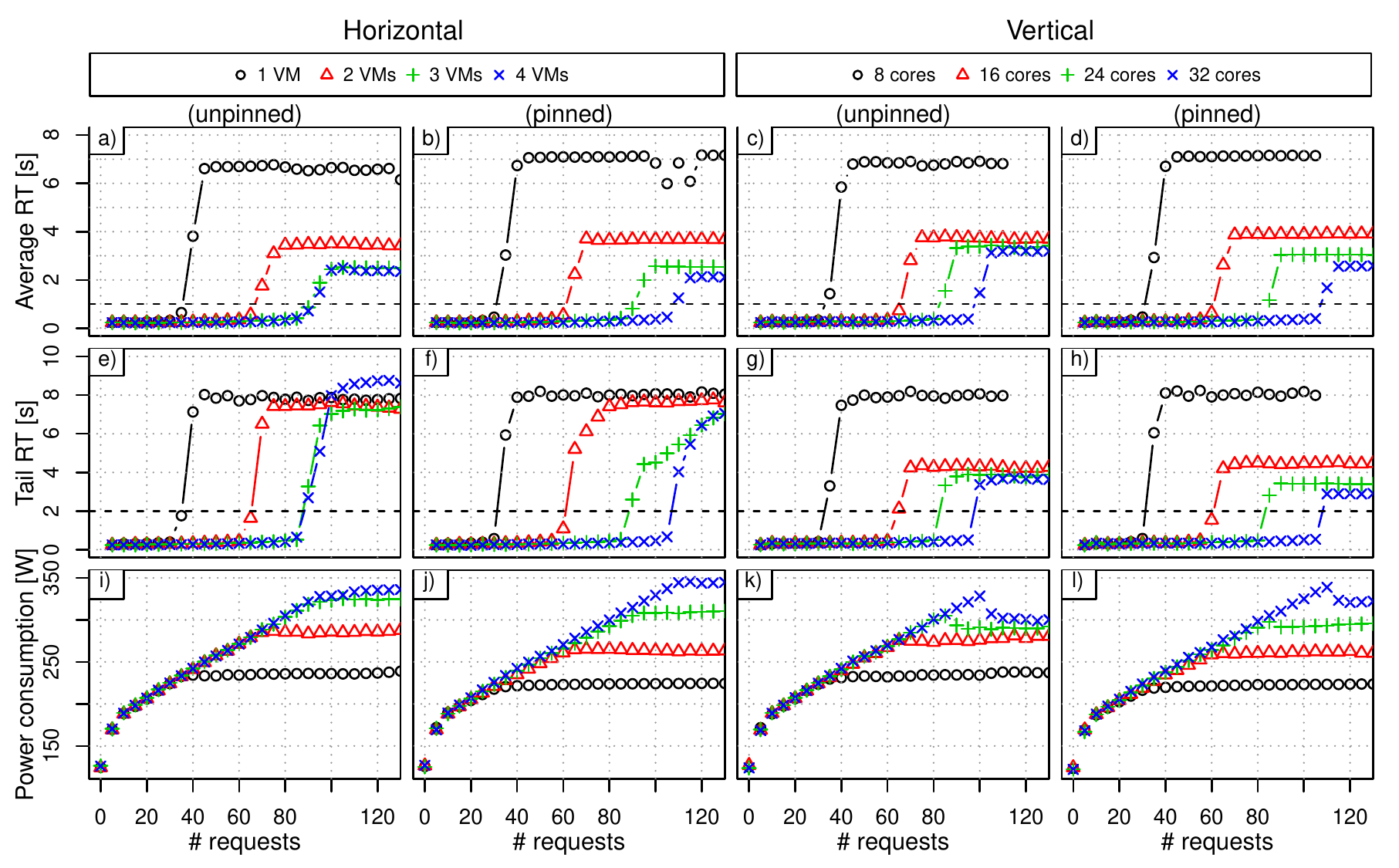}
    \caption{Comparison of horizontal and vertical scaling and the influence of CPU pinning on the response time (RT) and power consumption.}
    \label{fig:memcached_vertical_horizontal}
\end{figure*}

Figure~\ref{fig:memcached_vertical_horizontal} shows the influence of workload intensity (x-axis), type of scaling and CPU pinning (columns), and a size or a number of virtual machines (data series), on the average response time (the first row), tail response time (the second row), maximal throughput (the intersect of data series and the dashed horizontal line), and the power consumption (the third row). 

\subsection{Horizontal scaling}

In the horizontal scaling experiments we use virtual machines with 8 CPU cores and 10~GB of RAM.
We scale the number of virtual machines from 1 to 4 instances.
All instances are hosted on a single server to evaluate its power-performance tradeoffs.

\subsubsection{Influence of workload intensity}
First, we evaluate the relation between the number of virtual machines and the throughput of the application exposed to a step workload.
Scaling out (increasing the number of virtual machine instances) extends the maximum number of requests that the application is able to handle before becoming overloaded (Figure~\ref{fig:memcached_vertical_horizontal}a).
However, after reaching some number of virtual machines (3 virtual machines in our case, with 75\% of available CPU cores allocated) further scaling out is not beneficial, showing a non-linear relationship between the response time and allocated capacity~\cite{lakew2014towards}.
With a threshold on the average response time on 1~second
an application consisting of a single virtual machine can handle up to 35 requests,
two virtual machines allow to handle a workload up to 65 requests,
and an application consisting of three or four virtual machines can serve up to 90 requests.
Possible reasons for the lack of improvement when adding the fourth virtual machine are saturation of some resources or background jobs (e.g. operating system) that reduce the pool of available resources (adding the fourth virtual machine does not increase the pool of resources to 100\% of available).

Next, we investigate the effects of scaling out on the average and tail response times.
Horizontal scaling does not affect the response time when the application is underloaded (Figure~\ref{fig:memcached_vertical_horizontal}a).
When one or few instances can handle the workload, adding another instance does not improve the response time significantly.
For an overloaded application, increasing the number of virtual machines reduces the average response time, however that reduction is not proportional to the amount of resources added.
The average response time of an overloaded application stabilizes at a higher level when fewer virtual machines are used, because the application is overloaded to a higher extent.
For example, for a system with a single virtual machine, while the application is able to serve only 35 requests per second before overloading, there are up to 200 requests in the system (including delayed ones).
Changing the limit on the maximum number of requests present in the system has a significant impact on the average time of overloaded application.
For example, the average response time of a system with two virtual machine exposed to a constant workload of 100 requests per second,
varies from 1.6~s for a limit of 100 requests, through 3.2~s for 200 requests, to 4.7~s for 300 requests.

Moreover, scaling the system horizontally while using the round robin load balancing policy (a default one in HAProxy) does not improve the tail response time when the application is overloaded (Figure~\ref{fig:memcached_vertical_horizontal}e).

Finally, we analyze how the power consumption of a physical server changes depending on the number of virtual machines used for serving the workload.
Scaling in (decreasing the number of virtual machines) does not reduce the power consumption of a physical server when the application is underloaded (Figure~\ref{fig:memcached_vertical_horizontal}i).
The limited number of virtual machines influences only the maximal power consumption, when the application is overloaded.

\subsubsection{Influence of CPU pinning}
In the next experiment, we study the influence of CPU pinning on the effects of horizontal scaling.
Pinning virtual CPUs to the first $n$ physical CPU cores negatively impacts the performance of an application with one and two virtual machines, as shown in Figure~\ref{fig:memcached_vertical_horizontal}b,f.
For these configurations, the system is able to serve a lower number of requests under the same thresholds compared to the unpinned configurations (Figure~\ref{fig:memcached_vertical_horizontal}a,e).
On the other hand, pinning improves the performance of an application that uses all the CPU cores (four virtual machines in our case).
In this configuration the system can server up to 105 requests under the thresholds (compared to 90 requests, when virtual CPUs were not pinned).
CPU pinning decreases the power consumption of the system with one, two, and three virtual machines, while slightly increasing the power consumption of a system with four virtual machines (Figure~\ref{fig:memcached_vertical_horizontal}j).

\subsubsection{Influence of total number of requests}
\label{sec:pending-requests-analysis}
To understand why the tail response time is not improved by horizontal scaling we have performed additional experiments.

First, we investigate how the limit of the maximum number of requests present in the system influences the response time.
As we have mentioned in Section~\ref{sec:workload-generator}, we have limited the total number of not processed requests to 200 in order to study the behaviour of a saturated server.
That limit can be translated to the length of a queue at each virtual machine, e.g., since the application with a single virtual machine gets overloaded with a workload of 35 requests per second, the queue length is equal to 165.
Because of that limit, the response time of a saturated system stabilizes at some level before the application breaks
(i.e. a service becomes unavailable or an internal server error occurs).
In this experiment, we want to find out if that limit influences both the average and tail response time, and to what extent.

We expose the application to a constant workload that leads to
an overload
(10 requests per second more than the application is able to handle before getting overloaded).
Therefore, the exact number of requests generated per second depends on the number of virtual machines, and equals
45 requests for a single virtual machine,
75 requests for two virtual machines,
and 95 requests for three virtual machines (see Figure~\ref{fig:memcached_vertical_horizontal}a).
We repeat the measurements for various queue lengths (levels of overload): 0, 10, 25, 50, 75, and 100.

\begin{figure}[t]
    \centering
    \includegraphics[width=0.6\textwidth]{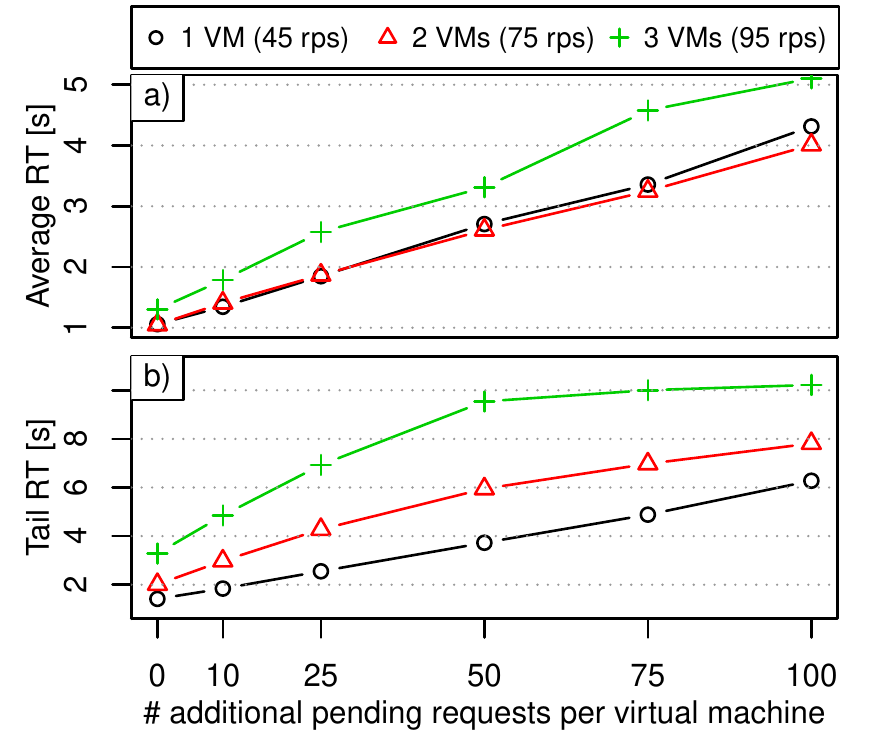}
    \caption{Influence of the total number of not processed requests on the response time (RT) of a horizontally scaled application.}
    \label{fig:scaling_workers}
\end{figure}

Figure~\ref{fig:scaling_workers} shows the effect of the total number of not processed requests present in the system on the average and tail response time.
Values on the x-axis are scaled by the number of virtual machines, and shifted in such a way, that 0 refers to the number of requests of a saturation workload.

The average response time grows linearly with the increased number of additional not processed requests (Figure~\ref{fig:scaling_workers}a).
Overlapping values for a system with one and two virtual machines show that the system scales almost perfectly at the beginning.
However, when the workload increases further, adding a third virtual machine does not allow to maintain the system performance.
A system with three virtual machines is not able to handle proportionally increased workload with the same average response time.
Also the tail response time changes with an increase of the number of not processed requests (Figure~\ref{fig:scaling_workers}b).
The tail response time scales worse than the average response time already from the beginning.
Even though, the system is exposed to the proportionally increased workload, scaling out does not allow to keep the tail response at the same level.

We have shown that \textbf{by changing the limit of the number of requests present in the system one can regulate to what extent the application is overloaded and by that influence both the average and tail response time}.
At the same time, the obtained results do not explain why the tail response time of an overloaded application is not improved by horizontal scaling.
Therefore, we continue our investigation with analysing other potential reasons.

\subsubsection{Influence of load balancing strategy}
Another element of the system
being evaluated
while trying to explain the lack of improvement in tail response time is the load balancer.
We have observed that when we use the default strategy for load balancing in HAProxy -- round robin -- some virtual machines got overloaded more than the other, and their performance was much worse.
That does not affect the average response time, since the other virtual machines still perform well, and the vast majority of requests is processed in short time.
However, the overload of even a single virtual machine has a big impact on the tail response time, because it affects enough requests to significantly increase the 95 percentile of response time. 
Therefore, we repeat the experiment for horizontal scaling with a different load balancing policy provided by HAProxy -- least connections -- that directs new requests to the server with the lowest number of connections.

\begin{figure}[t]
    \centering
    \includegraphics[width=0.6\textwidth]{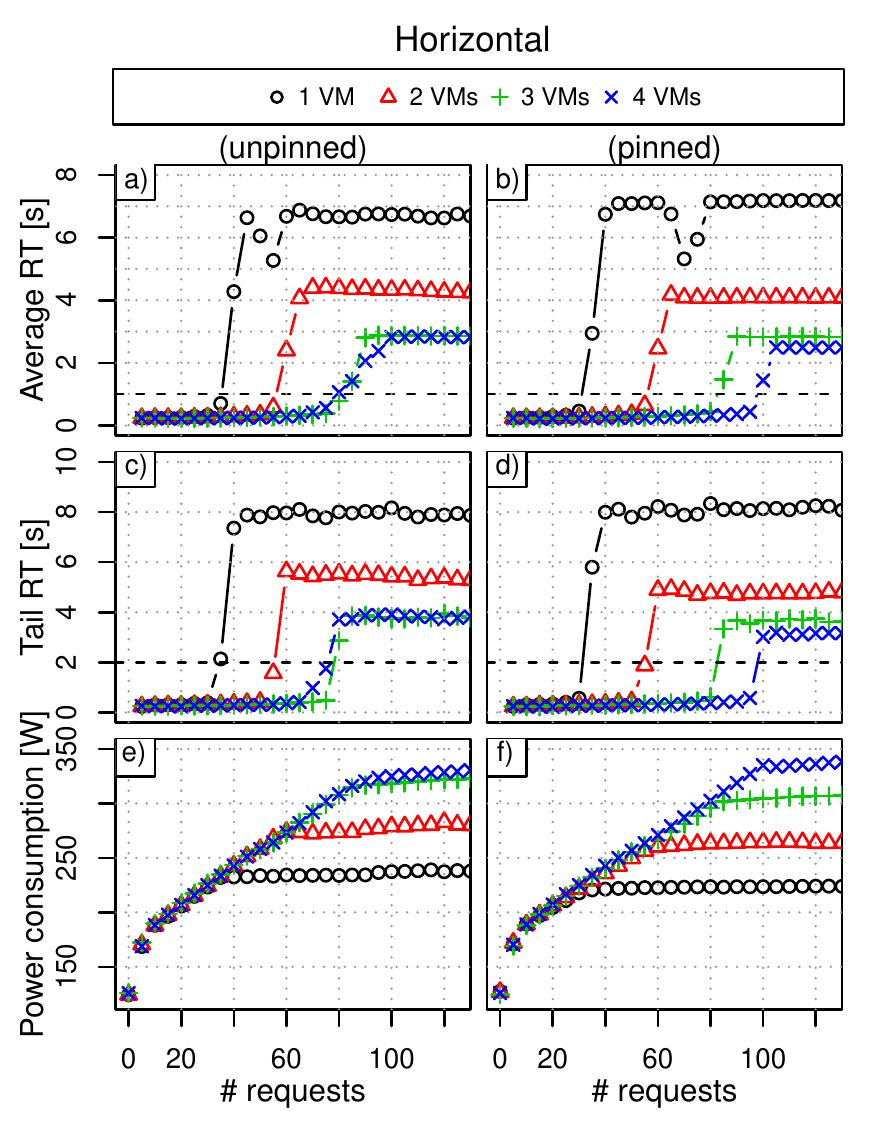}
    \caption{Influence of load balancing strategy on the response time (RT) and power consumption.}
    \label{fig:horizontal_scaling_load_balancer}
\end{figure}

Figure~\ref{fig:horizontal_scaling_load_balancer} shows the influence of workload intensity (x-axis), CPU pinning (columns), and a number of virtual machines (data series), on the average response time (the first row), tail response time (the second row), maximal throughput (the intersect of data series and the dashed horizontal line), and the power consumption (the third row). 

The change of the load balancing strategy results in a lower tail response time for configurations with two, three, and four virtual machines (Figure~\ref{fig:horizontal_scaling_load_balancer}c) compared to the configurations with the round robin load balancing strategy (Figure~\ref{fig:memcached_vertical_horizontal}e).
However, that improvement comes at a cost of a degraded average response time of an application with two, three, and four virtual machines (Figure~\ref{fig:horizontal_scaling_load_balancer}a).
The power consumption is not affected significantly by the change of the load balancer strategy (Figure~\ref{fig:horizontal_scaling_load_balancer}e).
Also the influence of CPU pinning (Figure~\ref{fig:horizontal_scaling_load_balancer}b,d,f) is very similar to what we have observed for the configurations with round robin load balancer strategy.

We conclude that \textbf{the load balancing strategy can drastically change the characteristic of the tail response time of a horizontally scaled application}.
Physical servers are complex heterogeneous systems, and the performance of virtual machines hosted on a single server may differ due to saturation of some shared components.
Therefore, load balancing strategies that monitor and adjust the performance of virtual machines to avoid overloads are needed to keep the tail response time low.
Based on evaluations of load balancing algorithms conducted by other researchers (e.g.,~\cite{durango2014control}), we have decided to use an algorithm that follows the shortest queue first principle.
Alternative approaches include to monitor the current performance of virtual machine (e.g., the response time) and direct the load to the virtual machines with the best performance (e.g., using HAProxy round robin with dynamically adjusted weights).
However, such approaches require well-tuned controllers for choosing appropriate weights, which is a complex problem that is outside the scope of our current work.

\subsection{Vertical scaling}

In vertical scaling experiments we use a single virtual machine and change both the number of the allocated CPU cores and the size of RAM.
We evaluate four configurations:
8 CPU cores with 10~GB of RAM,
16 cores with 20~GB,
24 cores with 30~GB,
and 32 cores with 40~GB.

\subsubsection{Influence of workload intensity}
First, we analyze the influence of the vertical scaling on the throughput.
Clearly, the smallest virtual machine (with 8 CPU cores and 10~GB of RAM) has worse performance than the bigger virtual machines.
A virtual machine with 8 cores is able to handle close to 30 concurrent requests with the average response time under 1~second (Figure~\ref{fig:memcached_vertical_horizontal}c).
By scaling up the virtual machine by 8 cores and 10~GB of RAM (up to 16 cores and 20~GB of RAM in total) one extends the ability of the application to handle the workload of up to 65 requests with the average response time under 1~second.
Adding another 8 cores and 10~GB of RAM (a virtual machine with 24 cores and 30~GB of RAM) results in an ability to serve up to 80 requests while not overloaded.
Finally, a virtual machine with 32 cores and 40~GB of RAM is able to handle up to 95 requests with an average response time under 1~second.

Second, we investigate how vertical scaling affects the response time.
Similarly to horizontal scaling, adding more resources reduces the average response time (Figure~\ref{fig:memcached_vertical_horizontal}c) and the the tail response time of an overloaded application (Figure~\ref{fig:memcached_vertical_horizontal}g).
While scaling up from 8 to 16 CPU cores reduces the response time by around 50\%, further scaling up does not give so significant improvements in the application performance.
Both the average and tail response time stabilize at the similar level for virtual machines with 16 CPU cores and more.

The power consumption, shown in Figure~\ref{fig:memcached_vertical_horizontal}k, differs significantly only for a virtual machine with 8 cores, when the application is overloaded (for a workload over 40 requests).


\subsubsection{Influence of CPU pinning}
In the final experiment, we investigate the consequences of CPU pinning when scaling a virtual machine hosting MediaWiki with Memcached vertically.
\textbf{CPU pinning improves the performance of very large virtual machines utilising most of the CPU cores of the host physical server} (24 and 32 cores in our experiments, Figure~\ref{fig:memcached_vertical_horizontal}d).
The largest virtual machine is able to serve up to 105 requests with an average response time under 1~second comparing to 95 requests for a setting without CPU pinning.
Also the average response time of an overloaded application decreases by approximately 1~second.

CPU pinning has a significant influence on the power consumption when an application is scaled vertically (Figure~\ref{fig:memcached_vertical_horizontal}l).
First, CPU pinning makes the power consumption of virtual machines with different sizes more differentiable.
While for not pinned virtual machines the difference in power consumption among virtual machines with 16 and more CPU cores allocated is smaller than 25~W (Figure~\ref{fig:memcached_vertical_horizontal}k), the difference between the 16 cores virtual machines and the 32 cores virtual machine is around 60~W when virtual CPUs are pinned.
Second, \textbf{CPU pinning decreases the power consumption of small and medium virtual machines that utilise a half or less of CPU cores of the host physical server (8 and 16 cores in our case), because the remaining part of CPU can enter idle states}.
Third, the power consumption of the virtual machine with 32 cores allocated is higher when virtual CPUs are pinned.

\section{Conclusions and Future Work}
\label{sec:conclusions}

In this paper we have analysed the power-performance tradeoffs of four techniques: DVFS, horizontal and vertical scaling, and CPU pinning.
We have performed a series of experiments on two types of physical servers (with AMD and Intel processors) using a real application -- MediaWiki.

\begin{table}[t]
\centering
\caption{Comparison of power-performance tradeoffs of all evaluated actuators} \label{tbl:comparison}
\begin{tabular}{rrr}
\hline
Actuator    & Power consumption reduction [\%] & Performance degradation [\%]\\ \hline
DVFS                & 14    & 33  \\
CPU Pinning         & 8     & 14  \\
Horizontal scaling  & 29    & 61  \\
Vertical scaling    & 27    & 68    
\end{tabular}
\end{table}

Table~\ref{tbl:comparison} compares the power-performance tradeoffs of all analysed actuators on a saturated AMD server.
The power consumption reduction is calculated as a difference between the power consumptions at the more power consuming setting (2.1~GHz, without CPU pinning, four virtual machines, or a virtual machine with 32 cores) and the less power consuming setting (1.4~GHz, with CPU pinning, one virtual machine, or a virtual machine with 8 cores) divided by the power consumptions at the more power consuming setting.
Similarly, the performance degradation is the difference in the maximal throughput with the average response time lower than 2~s between the above-mentioned settings divided by the higher value.
In general, a reduction in power consumption causes a performance degradation that is at least two times higher in the value.

Based on the obtained results, we give a set of recommendations for using DVFS, CPU pinning, and horizontal and vertical scaling taking into account the power-performance tradeoffs.
Table~\ref{tbl:conclusions} presents our suggestions on how the analysed actuators can be used to optimise the data center configuration.
We list various \emph{conditions} that describe the initial state of the system, together with recommended \emph{actions} and their \emph{influence} on the power consumption (\emph{PC}) and response time (\emph{RT}).

\begin{table}[t]
\scriptsize
\centering
\caption{Applicability of DVFS, CPU pinning, horizontal and vertical scaling for optimisation of physical server power consumption (PC) and application response time (RT)}
\label{tbl:conclusions}
\begin{tabularx}{\textwidth}{lXlll}
\hline
\textbf{Actuator}       & \textbf{Conditions} & \textbf{Actions}   & \textbf{Influence on PC}  & \textbf{Influence on RT}   \tabularnewline \hline

\multirow{3}{*}{DVFS}   & not saturated server and concurrent requests
                        & do nothing            & ---                       & ---               \\\cline{2-5}

                        & not saturated server and evenly spread requests
                        & reduce CPU frequency  & small positive            & small negative    \\\cline{2-5}

                        & saturated server                             
                        & reduce CPU frequency  & big positive              & big negative      \\
\hline
\multirow{3}{*}{CPU pinning}
                        & underloaded \newline application                            
                        & do nothing            & ---                       & ---               \\\cline{2-5}
                        
                        & overloaded application using a subset of CPU cores
                        & collocate virtual CPUs & positive                 & negative          \\\cline{2-5}
                                                                        
                        & overloaded application using all CPU cores
                        & pin virtual CPUs to reduce RT & negative          & positive          \\
\hline
\multirow{3}{*}{Horizontal scaling}
                        & underloaded \newline application                            
                        & do nothing            & ---                       & ---               \\\cline{2-5}
                        
                        & overloaded application using a subset of CPU cores
                        & scale out to reduces RT & negative                & positive          \\\cline{2-5}                   
                                                                        
                        & overloaded application
                        & scale in to save energy & positive                & negative          \\                  
\hline
\multirow{5}{*}{Vertical scaling}                       
                        & underloaded \newline application                            
                        & do nothing            & ---                       & ---               \\\cline{2-5}
                        
                        & overloaded application using a subset of CPU cores
                        & scale up to reduces RT & negative                 & positive          \\\cline{2-5}
                        
                        & overloaded application
                        & scale down to save energy     & positive          & negative          \\  
                                                                        
                                                                        
\end{tabularx}
\end{table}

The experimental results have shown that DVFS, CPU pinning, and virtual machine scaling have a significant influence on the power consumption only when the physical server is highly utilized.
There are two main reasons for that.
First, at the low utilization levels the total power consumption of a physical server is highly influenced by the idle power consumption part.
The above-mentioned techniques affect only the dynamic part of power consumption and therefore do not reduce significantly the total power consumption.
Second, when the system is not highly utilized, the CPU spends a lot of time in the idle states, when DVFS, CPU pinning and virtual machine scaling techniques simply do not influence the power consumption.

We conclude that the usefulness of DVFS, CPU pinning, and virtual machine scaling techniques for saving energy depends primarily on  the server utilization level.
The techniques are the most useful for data center operators to throttle overloaded applications in order to keep the power consumption of a physical server on an acceptable level (power capping).
However, the ability of these techniques to save energy at the low utilization levels is very limited.

The research presented in this paper may be extended in few directions.
First, the utility of DVFS, CPU pinning, and virtual machine scaling for the purpose of power capping could be further investigated. 
Towards this end, it would be useful to better understand how to throttle overloaded systems in the least obstructive way, i.e., to minimise application performance degradation while enforcing power caps. 
Second, the use of alternative actuators (e.g., Running Average Power Limit (RAPL) that allows control the power consumption of CPU sockets and DRAM~\cite{david2010rapl}) is planned to be evaluated against already established techniques for energy efficiency in data center server environments. 
Third, the power-performance tradeoffs of innovative approaches increasing the performance predictability~\cite{xue2016tale, dean2013tail} or reliability~\cite{sedaghat2016diehard, dong2013colo} of cloud computing systems should be assessed to understand at what cost the proposed improvements come.
Finally, we plan to investigate novel power and performance aware load-balancing algorithms.

\section*{Acknowledgement}

This work is funded by the Swedish Research Council (VR) project Cloud Control, the Swedish Government's strategic research project eSSENCE, the European Union's Seventh Framework Programme under grant agreement 610711 (CACTOS), and the European Union's Horizon 2020 research and innovation programme under grant agreement No. 732667 (RECAP).
Trevor E. Carlson is funded by the European Commission under the Seventh Framework Programme, grant agreement no. 610490.



\bibliographystyle{elsarticle-num}

\bibliography{sample}

\end{document}
